\documentclass{aa} 
\usepackage{graphicx}
\usepackage{subcaption}
\usepackage{txfonts}
\usepackage{url}
\usepackage[colorlinks=true,citecolor=blue,linkcolor=magenta,urlcolor=blue]{hyperref}
\usepackage{xcolor}
\usepackage{lscape}
\usepackage{booktabs} 

\newcommand{\msun}{${M}_{\odot}$}

\newcommand{\teff}{$T_\mathrm{eff}$}
\newcommand{\logg}{$\log g$}

\newcommand{\kms}{km\,s$^{-1}$}
\newcommand{\vsini}{$v_\mathrm{rot}\sin i$}

\begin{document} 

\title{Where are the binaries? - Searching for blue horizontal-branch stars in binary systems in the inner Galactic Halo}
\titlerunning{Where are the BHB binaries?}

\author{R.~Culpan \inst{1}
    \and M.~Dorsch \inst{1}
    \and I.~Pelisoli \inst{3}
    \and V.~Schaffenroth \inst{2}
    \and S.~Geier \inst{1}
    \and U.~Heber \inst{5}
    \and B.~Kub\'atov\'a \inst{4}
    \and H.~Dawson \inst{1}
    \and M.~Pritzkuleit \inst{1}
    \and A.~Bhat \inst{1}
    \and M.~Cabezas \inst{4}
    \and O.~Marjeva \inst{4}
    \and J.~Kub\'at \inst{4}
    \and M.~Kurpas \inst{1}
    \and J.~Vos \inst{1,4}
    \and F.~Mattig \inst{1}
    \and R.~Hainich \inst{1}}
\authorrunning{Culpan et al.}

\offprints{R.\,Culpan,\\ \email{rick@culpan.de}}

\institute{Institut f\"ur Physik und Astronomie, Universit\"at Potsdam, Haus 28, Karl-Liebknecht-Str. 24/25, 14476 Potsdam-Golm, Germany
\and Thüringer Landessternwarte Tautenburg, Sternwarte 5, 07778 Tautenburg, Germany
\and Department of Physics, University of Warwick, Coventry, CV4 7AL, UK
\and Astronomical Institute AS CR, Fri\v{c}ova 298, 251 65 Ond\v{r}ejov, Czech Republic
\and Dr. Remeis-Sternwarte \& ECAP, Astronomical Institute, FAU Erlangen-Nürnberg, Sternwartstr. 7, 96049 Bamberg, Germany}

\date{Received 27/08/2025 \ Accepted xx/xx/2025}

 
  \abstract
   {Blue horizontal-branch (BHB) stars are evolved low-mass objects that have completed their core hydrogen burning main-sequence (MS) stage and have lost significant mass during the red giant phase culminating in the helium flash. They are, hence, very old objects that can be used as markers in studying galactic structure and formation history. Their formation requires significant mass loss during the red-giant phase, but the role of stellar interactions in this process remains unclear. Knowing the fraction of BHBs that exist in binary or higher multiple systems where mass transfer may occur will enhance our understanding of their stellar evolution.}
   {We determine the fraction of BHBs in binary systems over a wide range of separations in the inner Galactic Halo to constrain mass-loss mechanisms and evolutionary pathways.}
   {Using a catalog of 22,336 BHB candidates from \textit{Gaia} DR3, we analysed radial velocity variations found in spectra (263 spectra of 89 targets) acquired using the Ond\v{r}ejov Echelle spectrograph attached to the Perek 2m telescope at the Astronomical Institute of the Czech Academy of Sciences together with archival spectra from the Ultraviolet and Visual Echelle Spectrograph (UVES) and Fiber-fed Extended Range Optical Spectrograph (FEROS). We searched for wide common proper motion pairs, binary candidates with enhanced astrometric noise, and binaries  with astrometric orbital solutions in \textit{Gaia} DR3. Archival light curves from \textit{Gaia} DR3 and the Zwicky Transient Facility (ZTF) were checked for binary induced variations. Synthetic SEDs and binary detection probabilities were modelled to account for selection effects.}
   {We find a binary fraction of <2.2\% ($1\sigma$ confidence), far lower than the rates for their main-sequence (MS) and red-giant branch (RGB) progenitors (30–50\%). This suggests that BHBs are either not descendants of binary systems, or that existing companions do not survive the BHB formation process.}
   {The negligible binary fraction implies single-star evolution could dominate BHB formation, contrasting with EHB stars where binarity is critical. Our results challenge models of mass loss on the RGB and highlight the need for alternative mechanisms.}

\keywords{stars: horizontal branch -- stars: binaries}

\maketitle
\section{Introduction \label{sec:intro}}

All horizontal branch (HB) stars have evolved from low-mass (${\sim}0.8\rm \,M_{\odot}$ to ${\sim}2.3\,\rm M_{\odot}$) main sequence stars 
that underwent a helium flash at the end of the red giant phase during which mass loss has occurred. The helium flash happens once the electron-degenerate helium core of the red giant reaches a temperature high enough for helium core burning to initiate, which occurs when the helium core reaches a mass of ${\simeq}0.5\rm \,M_{\odot}$.

The horizontal branch comprises several subcategories of stars. At the low-temperature end are red clump stars, with effective temperatures ($T_{\rm eff}$) in the range $4{,}400\,\mathrm{K} \lesssim T_{\rm eff} \lesssim 5{,}000\,\mathrm{K}$ \citep{galenne18}. These are followed by red HB stars (RHB), with $4{,}500\,\mathrm{K} \lesssim T_{\rm eff} \lesssim 6{,}000\,\mathrm{K}$ \citep{behr03, matteuzzi23}, and RR~Lyrae variables, with $5{,}900\,\mathrm{K} \lesssim T_{\rm eff} \lesssim 7{,}200\,\mathrm{K}$ \citep{bono95}. Blue HB (BHB) stars span the range $7{,}000\,\mathrm{K} \lesssim T_{\rm eff} \lesssim 20{,}000\,\mathrm{K}$ \citep{catelan09}, while extreme HB (EHB) stars -- also referred to as hot subdwarfs -- have $T_{\rm eff} \gtrsim 20{,}000\,\mathrm{K}$ \citep{heber09, heber16}. The range of effective temperatures seen along the HB is due to the range of masses of the hydrogen envelope around the helium burning core. The corresponding hydrogen envelope masses range from $\gtrsim$0.1\,$M_{\odot}$ for red clump stars \citep{girardi16} to $\lesssim 0.01\,M_{\odot}$ for hot subdwarfs \citep{heber16}.

The different hydrogen envelope masses of HB stars arises from different amounts of mass loss on the red giant branch (RGB). There are many mass-loss mechanisms at work during the red-giant phase. Mass loss rates depend on factors such as metallicity \citep{mcdonald15}, age \citep{metcalfe22}, rotation \citep{blackman16,shoda20, gehan22}, binarity \citep{vanrensbergen20}, magnetic fields \citep{gehan22, vanrensbergen20}, and binarity \citep{moe19}. Binary interactions can strip hydrogen envelopes via Roche-lobe overflow (RLOF), but their role in BHB formation is unclear.

Metallicity not only affects mass loss, but plays a key role in stellar evolution as a whole. Metal poor solar-mass stars have cores that are smaller and hotter than higher metallicity stars. This, in turn, will make low metallicity stars leave the MS quicker than high metallicity stars as the higher temperature increases the efficiency of hydrogen burning \citep{byrne25}. The RGB phase will also take longer in higher metallicity stars due to their cooler, more extended envelopes and delayed helium ignition \citep{karakas22}. Lower metallicity stars require a lower core mass for helium ignition than higher metallicity stars. The higher metallicity stars will, however, lose more mass as metals are critical for driving stellar winds through line absorption and have increased opacities in their atmospheres enhancing the effectiveness of radiation-driven mass loss. There is a strong interdependence between rotation, binarity and magnetic fields \citep{moe19}. Magnetic fields create a coupling between the red giant and its stellar wind resulting in the loss of angular momentum and the mass loss. Red giants that are in close binary systems ($P \, {\gtrsim} \, 150$ days) spin faster than single red-giants resulting in a stronger magnetic field \citep{gehan22}.

\citet{moe19} studied the binary fractions of MS stars and giants, the progenitors of BHB stars, as a function of metallicity including data of several large surveys and correcting them for selection biases. They found a strong anti-correlation of close binary ($P<10^{4}\,{\rm d}$, $a<10\,{\rm au}$) fraction with metallicity. Low-metallicity stars have significantly higher close binary fractions with on average shorter orbital periods than solar metallicity stars. The fractions for metallicities [Fe/H]=-1.0 and [Fe/H]=-3.0 are $40\pm6\%$ and $53\pm12\%$, respectively. They are very similar for dwarfs and giants. The fraction of wide binaries ($<200\,{\rm au}$) was found to be unaffected by metallicity.

In order to understand whether binarity is of importance as a mass-loss mechanism during the red giant phase for BHB progenitors we also need to examine the rates of binarity observed for other stars on the horizontal branch as well as the progenitors of BHBs. 

The majority of hot subdwarfs are the helium cores of red giants that have had their outer hydrogen envelope stripped away through mass-loss at the tip of the red giant branch, most likely driven by binary interactions \citep{geier22,schaffenroth22,pelisoli20, heber18}. They make up the extreme horizontal branch (EHB) at the bluest, hottest end of the horizontal branch. Two thirds of the hot subdwarfs in the young Galactic disk reside in post-interaction binaries, one third in very close post-common envelope (CE) systems. \citet{geier24} studied the radial-velocity variability of hot subdwarfs in the Galactic halo and found no post-CE systems at all and only some indications for binaries formed after stable mass transfer. Old globular clusters were found to be completely devoid of hot subdwarf binaries \citep{latour18}.

On the redder, cooler side of the BHB region of the CMD we find RR\,Lyrae stars on the instability strip. Metal-poor RR\,Lyrae stars in the Galactic Halo can be formed as single stars through stellar wind driven mass-loss on the red giant branch \citep{bobrick24}. \citet{bobrick24} investigated the population of metal-rich RR\,Lyrae stars in the \textit{Gaia} DR2 catalogue \citep{iorio21}. Their modelling showed that such stars cannot be explained by standard single-star evolution, as they are too young (1--9 Gyr) and too metal-rich ($\mathrm{[Fe/H]} \gtrsim -1$) to have reached the horizontal branch without additional mass loss. They concluded that all metal-rich RR\,Lyrae must have undergone binary interaction, where enhanced mass loss on the RGB enabled their evolution to the instability strip within their present age. In contrast, they predicted that halo RR\,Lyrae may host binary companions, but these would typically be in wide, non-interacting systems. Observationally, however, the confirmed binary fraction among RR\,Lyrae remains extremely small, with only two systems securely identified and up to $\sim$500 additional, but unconfirmed, candidates. This corresponds to an observed binary fraction of roughly $2$--$500$ relative to the total number of stars in the \textit{Gaia} RR\,Lyrae catalogue, i.e. at most on the order of $\lesssim 1\%$, highlighting the current uncertainty in constraining the true binary fraction in this population.

There are no precise, RHB-only binary-fraction measurements in the literature. Existing RV work \citep{carney08} that includes RHB stars is sample-limited and does not separate out a reliable RHB binarity fraction. \citet{matteuzzi23} states that the presence of solar metallicity ([Fe/H] $\approx$ 0) red HB stars requires a much higher mass-loss rate on the RGB than predicted by standard mass-loss models. Binarity would be a possible mass-loss enhancing mechanism that might explain this but published results supporting this could not be found.

\citet{belokurov20} analysed unresolved companions with \textit{Gaia} DR2. They found that both BHBs and blue stragglers showed higher astrometric uncertainties than photometrically similar objects, indicating higher binary fractions. The findings of \citet{belokurov20} are in contrast to the results of \citet{monibidin11} who studied close binaries in NGC 2808 and found 5\% - 10\% binarity for BHB stars with $12,000 K < T_{\rm eff} < 17,000K$ and periods <200 days.

\begin{table*}
  \centering
  \caption{Published binary fractions for horizontal-branch stars.}
  \label{tab:binary_rates}
  \begin{tabular*}{\textwidth}{@{\extracolsep{\fill}}lcccc@{}}
    \toprule
    \toprule
    Population & Binary Fraction & Method & Metallicity [Fe/H] & Reference \\
    \midrule
    BHB (this work) & $<2.2\%$ & RV + SEDs & $-1.7 \pm 0.3$ & -- \\
    EHB (field) & $\sim50\%$ & RV + light curves & $-1.5$ to $+0.2$ & \citet{geier22} \\
    RR\,Lyrae (Halo) & $<3\%^{*}$ & \textit{Gaia} + RV & $-1.5$ to $+0.2$ & \citet{bobrick24} \\ 
    RHB & undetermined & - & - & - \\    
    Red Clump (field) & $\sim20\%$ & APOGEE RV & $-0.5$ to $+0.5$ & \citet{badenes18} \\
    \bottomrule
  \end{tabular*}
  \vspace{2mm}
  \small * Based on 500 candidates from 17,570 catalogue  RR\,Lyr with full light curve information.
\end{table*}

Here we study the binarity of field BHB stars. The contents of this paper are as follows: In Section \ref{sect:spectroscopic} we present our spectroscopic analysis of 89 BHB candidates, including the methodology for radial velocity measurements, BHB/MS differentiation using rotational velocities and synthetic SEDs with \textit{Gaia} DR3 \verb!parallax! and our binary detection probability calculations. 
Section \ref{sect:photometric} details our photometric search for binaries using \textit{Gaia} DR3  and ZTF data, including the identification and analysis of variable sources. Section \ref{sect:astrometric} examines astrometric binary detection methods, focusing on \textit{Gaia} DR3's renormalised unit weight error (\verb!ruwe!) parameter where we re-evaluate the method used by \citet{belokurov20} and search for common proper motion pairs. 
Finally, Section \ref{sect:summary} summarises our findings, discusses their implications for stellar evolution and mass-loss mechanisms, and outlines directions for future research.

\section{Spectroscopic analysis}
\label{sect:spectroscopic}

The steps taken to calculate an upper limit on BHB binarity using time-resolved spectroscopy were: target selection, observation, data analysis, detailed classification, and the estimation of the binary detection probabilities.

We selected the BHB candidates to be observed from the catalogue of \citet{culpan21} and later refined the selection using \citet{culpan24} (henceforth C24). Both catalogues were generated using \textit{Gaia} DR3 CMD and tangential velocity cutoffs as the primary criteria. The CMD cutoffs used in \citet{culpan21} were revised using the Ond\v{r}ejov dataset in order to produce the improved C24 catalogue. 

Targets selected from these catalogues were limited to those with a \textit{Gaia} DR3 apparent $G$ magnitude of less than 11\,mag and an elevation of at least 30 degrees above the horizon at the Ond\v{r}ejov location at some time during the year. These selection criteria gave us a final list of 89 BHB candidates that were suitable for observation in Ond\v{r}ejov. A complete list of all the BHB candidates in the target list can be found in Appendix A.

\subsection{Spectroscopic dataset}

We acquired 263 spectra using the Ond\v{r}ejov Echelle Spectrograph \citep[OES,][]{koubsky04} attached to the Perek 2m telescope at the Astronomical Institute of the Czech Academy of Sciences which has a resolving power of $R$ = 51,600 around H$\alpha$.  Many of the spectra were acquired during the 2021, 2022, and 2023 workshops on observational techniques held jointly between the Stellar Physics Department at Astronomical Institute of the Czech Academy of Sciences and the Stellar Astrophysics Group at Potsdam University. Multiple spectra were acquired for all BHB candidate objects in order to be able to find radial velocity variations as an indicator of binarity.

\begin{figure}
  \centering
  \includegraphics[width=\hsize]{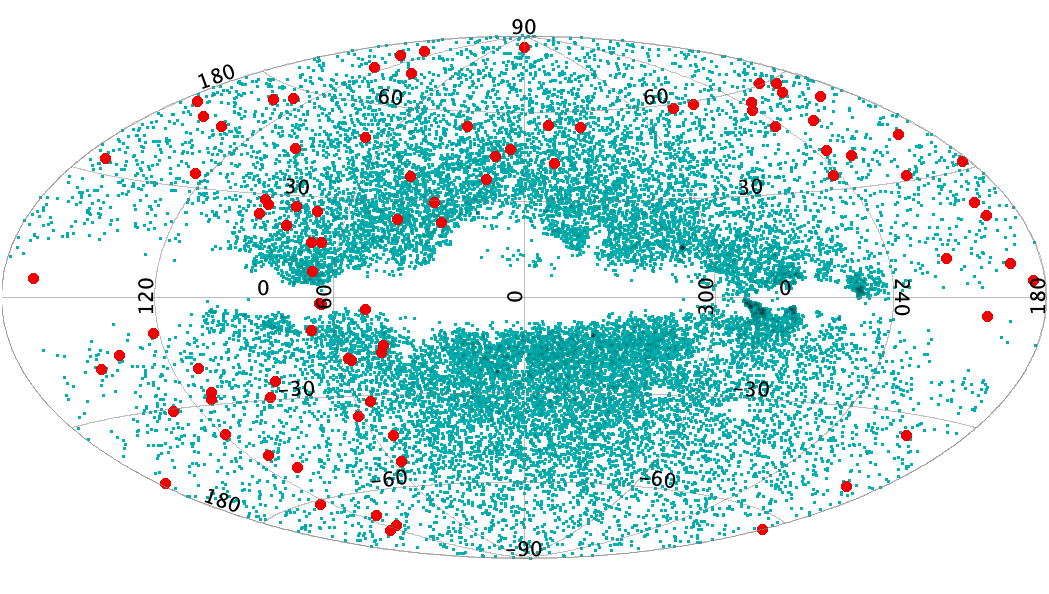}
 \caption{C24 BHB Candidates (cyan). BHB candidates that conformed to the Ond\v{r}ejov target list selection criteria (red).}
  \label{Ond\v{r}ejov_Sky}
  \end{figure}

The resulting OES spectra were reduced using the Image Reduction and Analysis Facility (IRAF)\footnote{IRAF is distributed by the National Optical Astronomy Observatories, which are operated by the Association of Universities for Research in Astronomy, Inc., under cooperative agreement with the National Science Foundation.} \citep{tody86,tody93} package and a dedicated semi-automatic pipeline \citep{cabezas23}. This includes all standard procedures for Echelle spectra reduction, including bias correction, flat-fielding, wavelength calibration, barycentric corrections, and continuum normalization. The acquired OES spectra had a mean signal-to-noise ratio ($\overline{SNR}$) of 30.2 and a standard deviation ($\sigma_{SNR}$) of 17.6.

We found that 23 of the BHB candidates with spectra acquired in Ond\v{r}ejov also had Ultraviolet and Visual Echelle Spectrograph (UVES) \citep{dekker00} or Fiber-fed Extended Range Optical Spectrograph (FEROS) \citep{kaufer99} spectra archival data. UVES, mounted on the ESO Very Large Telescope (VLT), covers a broad wavelength range, typically split into two settings: the blue arm (300–500 nm) and the red arm (420–1100 nm), with a spectral resolution of R $\approx$ 40,000–110,000 depending on the slit width and configuration. FEROS, installed on the ESO/MPG 2.2-m telescope, offers a single-shot wavelength coverage of 350–920 nm at a resolution of R $\approx$ 48,000.

The downloaded archival spectra from UVES and FEROS were already reduced. The radial velocity measurements for the UVES spectra had barycentric corrections applied by using a Python programme written specifically for this purpose. The reduced FEROS spectra already had this correction applied. Adding this 56 spectra to the 263 spectra acquired in Ond\v{r}ejov gave a total of 319.

\subsection{Spectral data analysis}

All 319 spectra were interpreted manually using the SPAS (Spectra Plotting and Analysis Suite). This is a specialised software tool developed in the University of Erlangen-Nuremberg, Germany for the visualisation, processing, and analysis of spectroscopic data in astrophysics. The software provides interactive plotting, spectral line fitting and equivalent width measurements \citep{hirsch09}. The grids of model spectra that were used for the spectral fitting were computed specifically for BHB stars using a hybrid LTE/NLTE approach as described in \citet{latour23} and references therein. We examined the H$\alpha$, H$\beta$, and H$\gamma$ spectral absorption lines to determine the rotational broadening by evaluating the width of the spectral lines and the radial velocity with regard to the observer by evaluating the Doppler shift of the centre of the spectral line with regard to the rest-frame wavelength. The H$\alpha$, H$\beta$, and H$\gamma$ lines were chosen for the interpretation of the spectra as only these prominent features were suitable given the varying data quality.

The individual RV measurements are provided in Table A.1 together with the weighted means of both the projected rotational velocities ($v_{\rm rot}\sin{i}$) and RVs. The RV results using the UVES and FEROS spectra matched those obtained from OES data demonstrating a high level of consistency between instruments.

The average statistical uncertainty of our RV measurements is $2.16\,{\rm km\,s^{-1}}$ which is consistent with the findings of \citet{gajdos24}. To check for systematic shifts we compared our mean RVs with independent high-precision measurements of 14 well-known BHB stars from the literature \citep{kinman00,behr03,kafando25}, which are part of our sample (BD+00\,145, BD+25\,2602, BD+42\,2309, HD\,2857, HD\,8376, HD\,60778, HD\,74721, HD\,86986, HD\,87047, HD\,87112, HD\,93329, HD\,109995, HD\,167105, HD\,203563, HD\,252940). The mean deviation is just $\pm0.9\,{\rm km\,s^{-1}}$ and the individual deviations are always smaller than the statistical uncertainties.

\subsection{BHB/MS differentiation}

As outlined in C24, the main contaminants in the BHB candidate catalogues are early MS A- and B-type stars. To make sure that we only select BHB stars for this study, we further refined the cleaning procedure described in C24. Due to the limited quality of the spectra, it was not possible to use the strengths of weak metal or helium lines to distinguish both types. Instead we used the \vsini, which are on average higher for MS-A/B stars, and a combination of SEDs and stellar evolution tracks.

\begin{figure}
  \centering
  \includegraphics[width=\hsize]{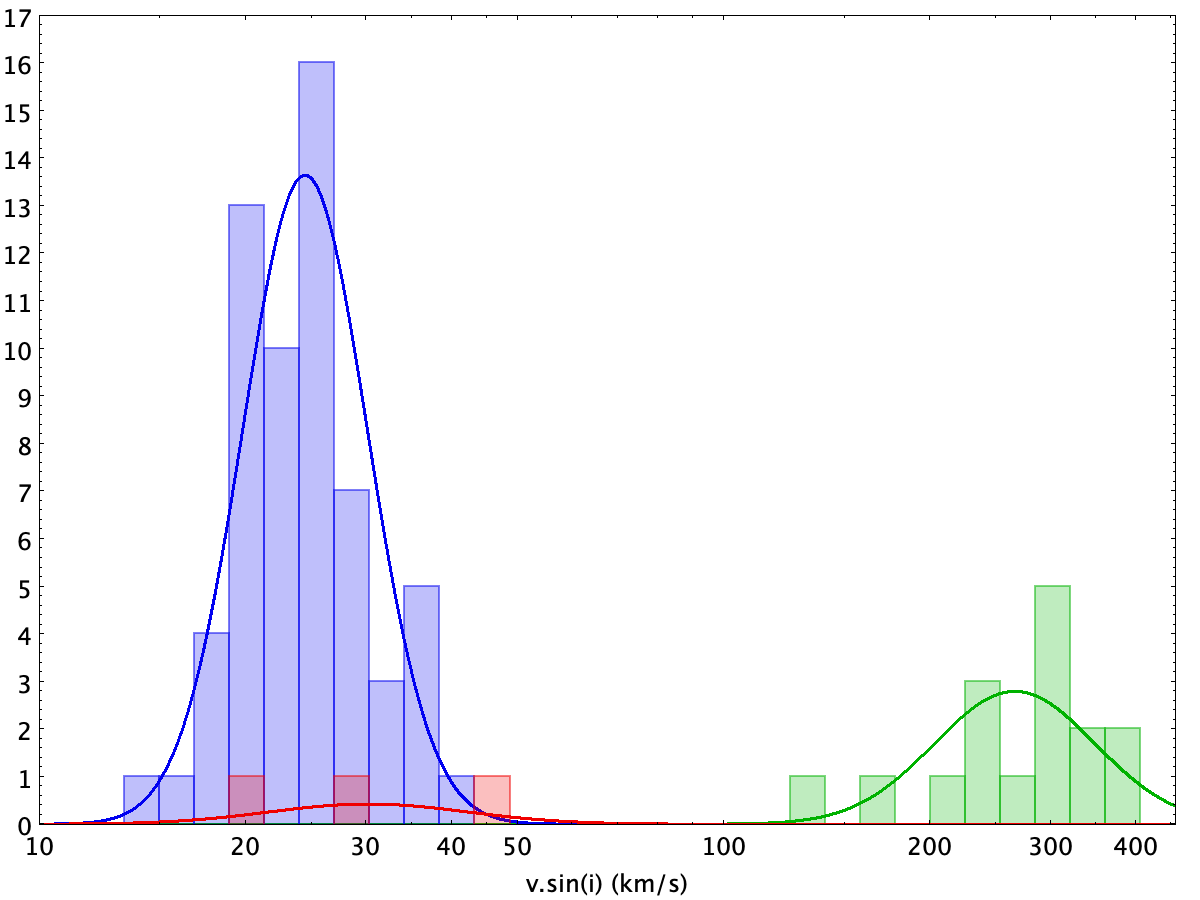}
 \caption{Targets confirmed as BHBs using Ond\v{r}ejov spectra and SEDs (blue), targets with low rotational velocity but confirmed as main sequence stars using SED (red), and targets confirmed as MS stars using SEDs and Ond\v{r}ejov spectra (green).}
  \label{V_sin_i_Ondrejov}
  \end{figure}

In order to distinguish between BHBs and main-sequence A- and B-type stars we initially used a \vsini\ cutoff where the candidates with \vsini\ < 50\,\kms were considered BHBs. This was based on the bimodal distribution of \vsini\ where 16 of the 89 C24 BHB candidates observed had a \vsini\ > 50\,\kms (see Figure~\ref{V_sin_i_Ondrejov}). The cutoff used is in line with the findings of \citet{behr03} who states that most BHB stars rotate with \vsini\ < 15 \kms, but with some "fast-rotators" with \vsini\ < 30-35 \kms.

In order to confirm our BHB/MS differentiation from \vsini\ we generated synthetic SEDs for each of the BHB candidates observed in Ond\v{r}ejov.  Here we used grids of synthetic SEDs calculated using a revision to the ATLAS12 \citep{kurucz96,irrgang18} code. More detail surrounding the SED fitting method can be found in \citet{heber18} and \citet{latour23}. 
We generated multiple synthetic SEDs over a range of effective temperatures (7,000\,K < $T_\mathrm{eff}$ < 15,000\,K) and with surface gravities (2.5 < \logg\ < 4.0) selected at zero-age, mid-age, and terminal age values for horizontal-branch stars, using Bag of Stellar Tracks and Isochrones \citep[BaSTI,][]{pietrinferni04,pietrinferni06,hidalgo18} HB tracks for $\mathrm{[Fe/H]} = -1.7$ based upon the metallicity of halo field stars as determined by \citet{jofre11}. 

Similarly, multiple SED fits were also performed using zero-age, mid-age, and terminal age values for main sequence objects and assuming [Fe/H] = $-$0.1.

The stellar radius was computed as $R = \Theta / (2\varpi)$, where $\Theta$ is the angular diameter and $\varpi$ is the \textit{Gaia} DR3 parallax corrected for the global zero point \citep{lindegren20}. The stellar mass was derived via $M = gR^2 / G$, and the luminosity was calculated from $L / L_\odot = (T_{\rm eff} / T_{\rm eff,\odot})^4 (R / R_\odot)^2$. Interstellar extinction was accounted for using the extinction law of \citet{fitzpatrick19} for a reddening parameter of $R_{55}=3.02$, an average value for the Galaxy. The quality of the fit was evaluated by finding the \teff\ and $\log g$ with the best $\chi^2$ fit within the BaSTI modelled isochrones \citep{pietrinferni04} for MS and HB stars respectively.

The values of $T_\mathrm{eff}$ and $\log g$ that gave the best fit between radius, mass, and luminosity when compared to the values expected from BaSTI tracks were considered to be the best solution. This was done using MS and BHB parameters. The results from the MS best-fit SED and the BHB best-fit SED were then compared in terms of standard error of the fit (the lower the better), the number of photometric data points used (the higher the better), and the reddening from extinction (as close as possible to published values) to determine whether MS or BHB parameters offered the best fit. 

In this way we have used synthetic SEDs to differentiate between BHBs and MS stars. The metallicity assumptions made were considered fit for purpose, as varying this figure did not change the category that was found. 

\begin{figure*}[t]
    \centering   
    \begin{subfigure}[t]{0.49\textwidth}
        \centering
        \includegraphics[width=1\textwidth]{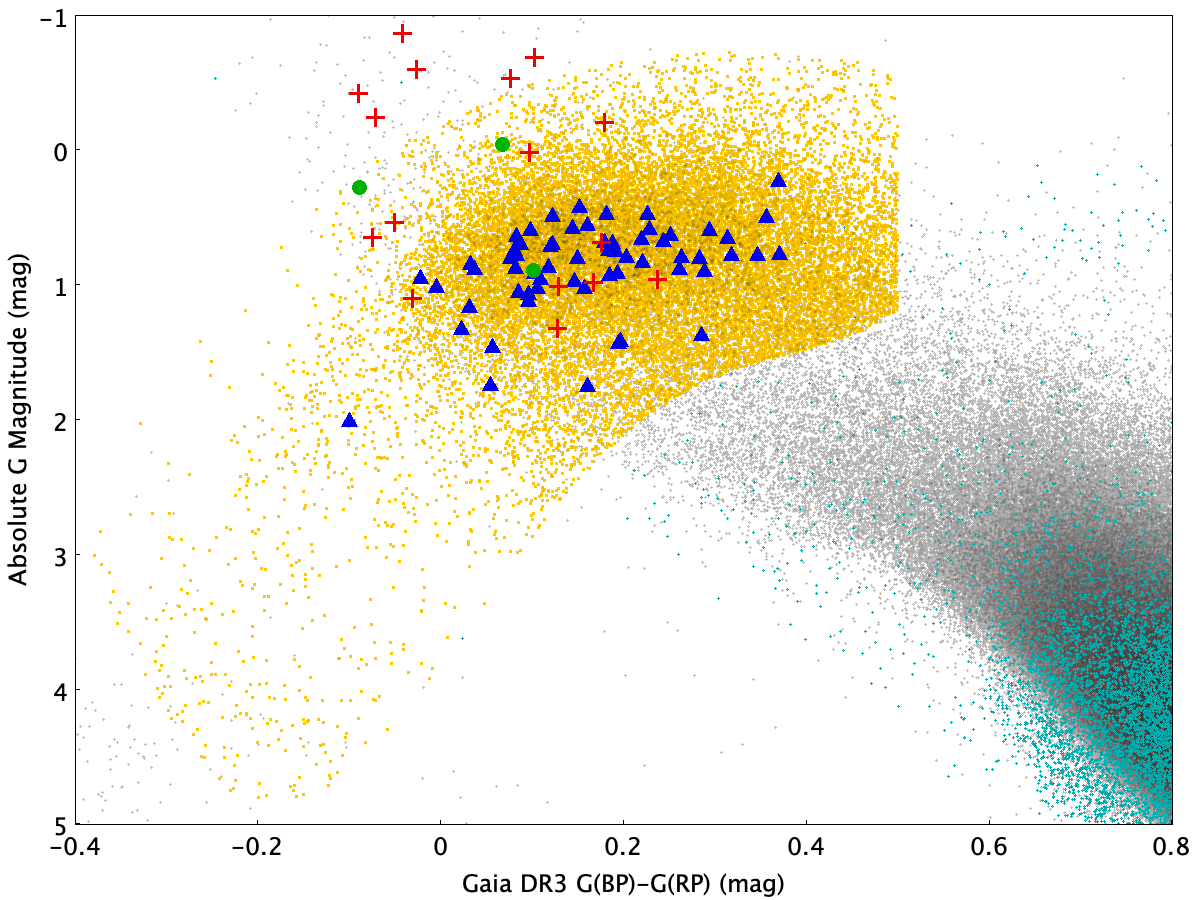}
    \end{subfigure}
    \hfill
    \begin{subfigure}[t]{0.49\textwidth}
        \centering
        \includegraphics[width=1\textwidth]{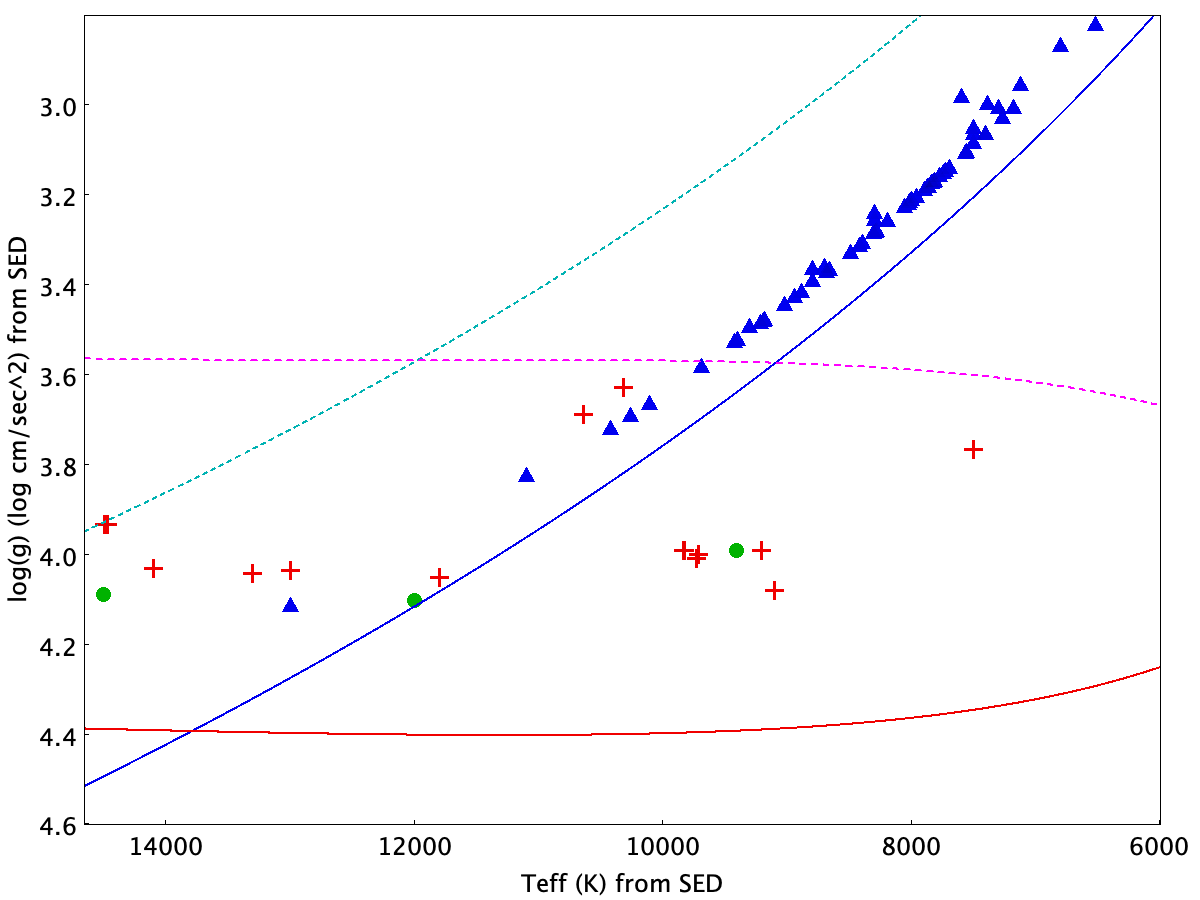}
    \end{subfigure}
    \begin{subfigure}[t]{0.32\textwidth}
        \centering
        \includegraphics[width=1\textwidth]{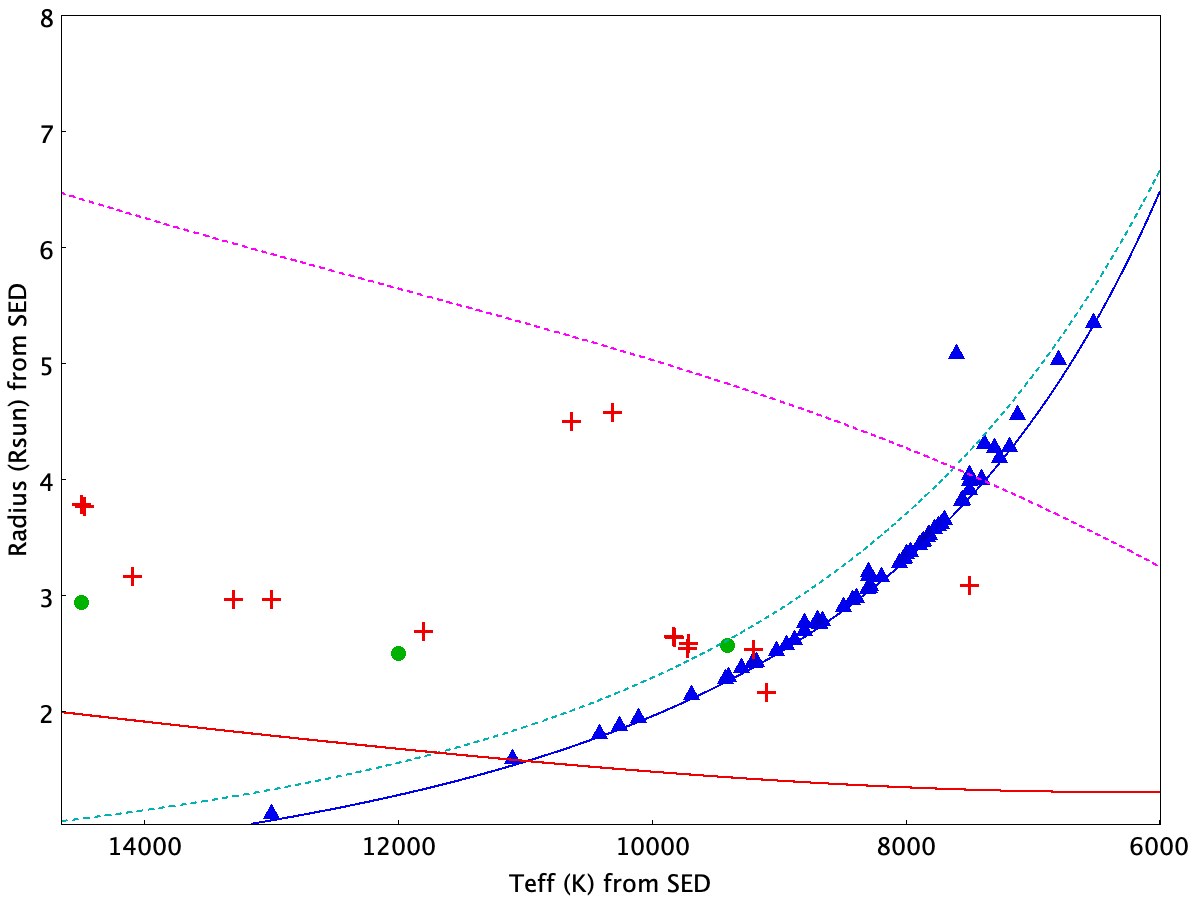}
    \end{subfigure}
    \hfill
    \begin{subfigure}[t]{0.32\textwidth}
        \centering
        \includegraphics[width=1\textwidth]{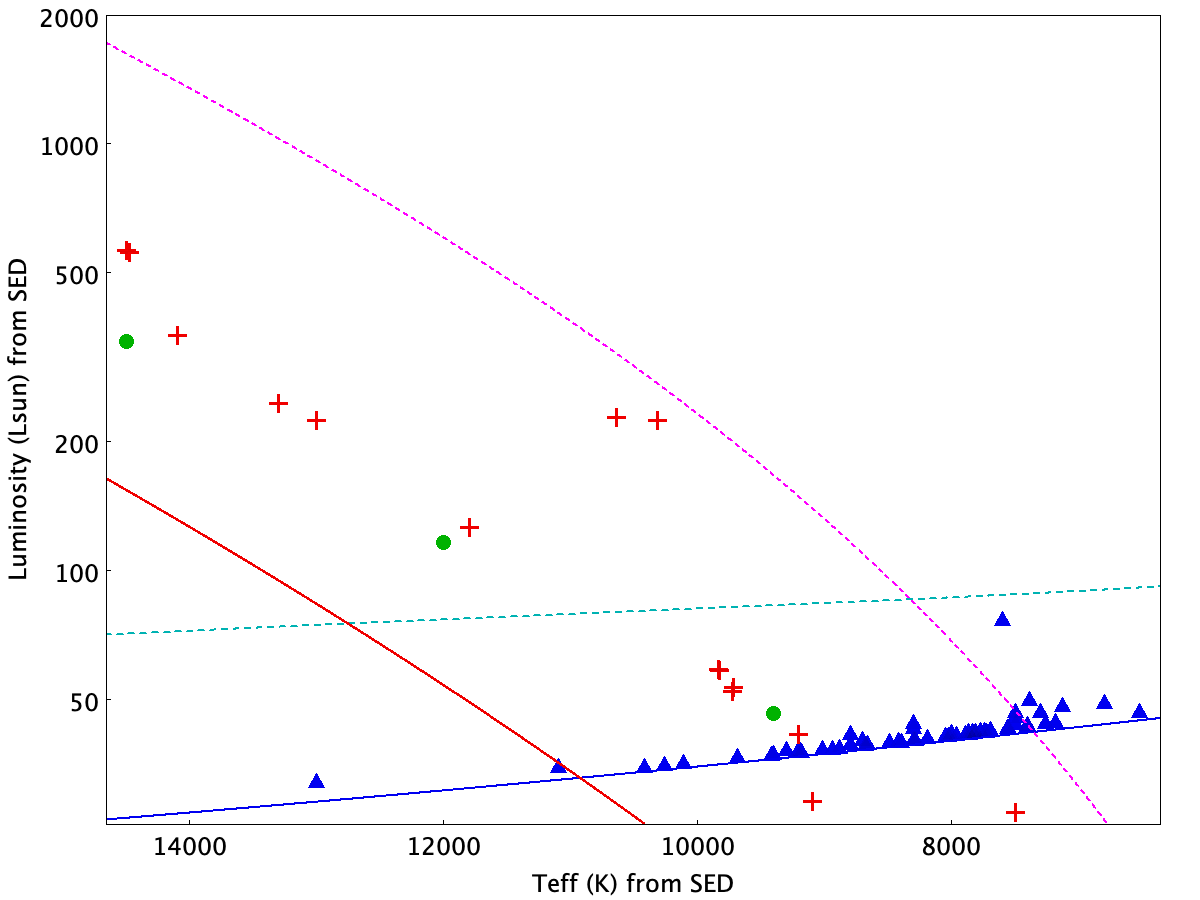}
    \end{subfigure}
    \hfill
    \begin{subfigure}[t]{0.32\textwidth}
        \centering
        \includegraphics[width=1\textwidth]{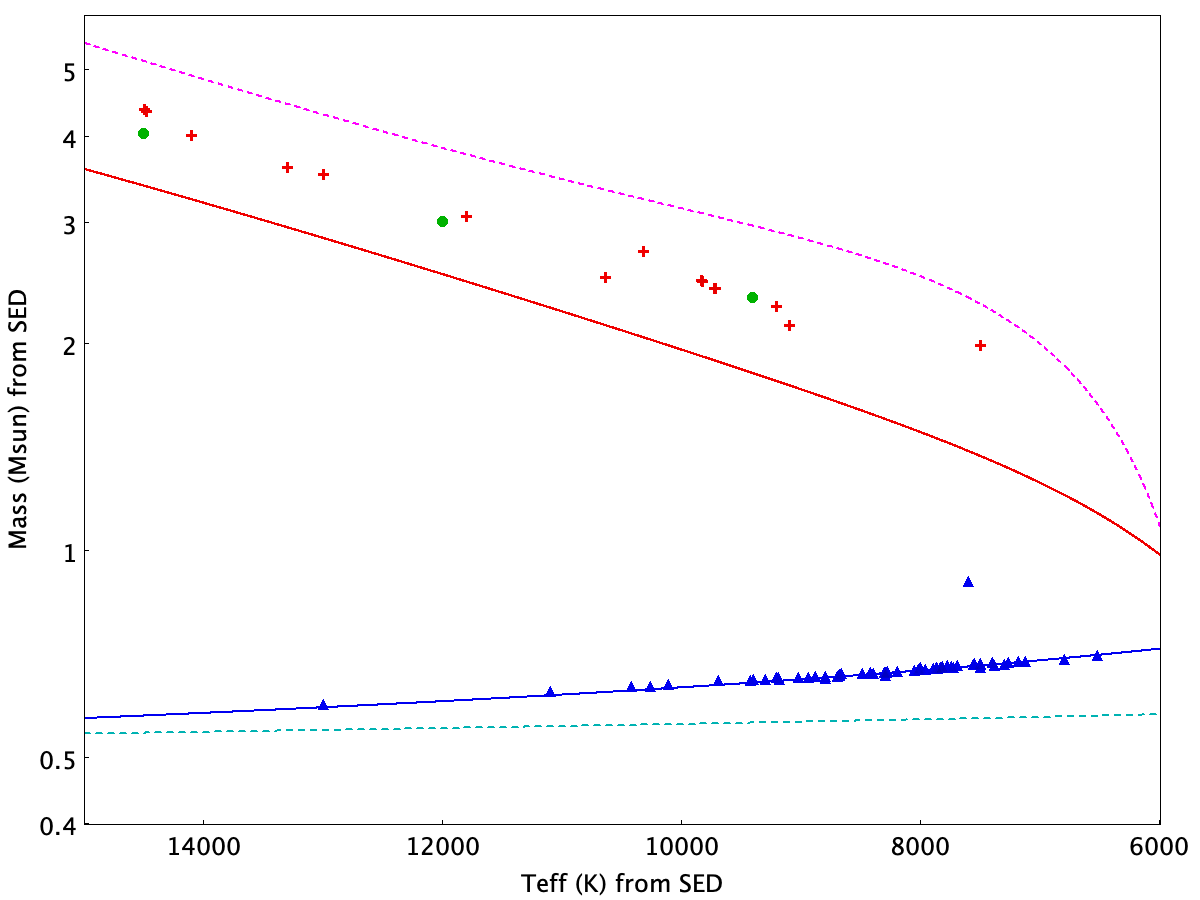}
    \end{subfigure}
     \caption{Upper left panel: \textit{Gaia} DR3 CMD for 30,000 randomly selected objects with parallax errors less than 10\% (grey dots), and the subset with tangential velocities greater than 145 \kms\ typical for halo stars (cyan dots). C24 BHB candidates (yellow dots). Upper right panel: $T_{\rm eff} -  \log g$ (Kiel) plot; Bottom left panel: $T_{\rm eff} -  R$ plot; Bottom middle panel: $T_{\rm eff} -  L$  plot; Bottom right panel: $T_{\rm eff} -  M$ plot for the 89 BHB candidates listed in Appendix A. All panels: Targets confirmed as BHBs using Ond\v{r}ejov spectra and SEDs (blue triangles), and targets confirmed as main sequence stars using Ond\v{r}ejov spectra and SEDs (red crosses), targets with low rotational velocities from Ond\v{r}ejov spectra but with MS stellar and atmospheric parameters from SED evaluation (green circles).}
  \label{Ondrejov_SED}
\end{figure*}

The synthetic SEDs found three BHB candidates with low rotational velocities that plot in the main-sequence regions of the $T_\mathrm{eff} -  \log~g$, $T_\mathrm{eff} -  R$, $T_\mathrm{eff} -  L$, and $T_\mathrm{eff} -  M$ plots (see Figure~\ref{Ondrejov_SED}). We consider the low measured value of rotational velocity to be due to the rotational axis of the fast-spinning main-sequence star being oriented towards the observer. Having 3 out of 19 MS contaminants showing \vsini\ values below the cutoff is in line with expectations for MS stars with random axial orientations. The classifications of the sample stars are provided in Table A.1.

\subsection{Binary detection probability}

All of the 89 stars had between 2 and 5 OES spectra acquired over a 35 month period from 5th May 2021 to 8th April 2024. Each target had spectra acquired on random dates throughout the period in question but conforming to the criteria given in Section 2.1. 23 of these stars also had UVES or FEROS spectra which were incorporated into our analysis.

As each target had spectra acquired with different time intervals between observations with varying observing conditions we had a heterogeneous data set where each of the spectra had differing signal-to-noise ratios, resulting in differing standard errors on the radial velocity measurements.

\begin{figure*}
  \centering
  \includegraphics[width=\hsize]{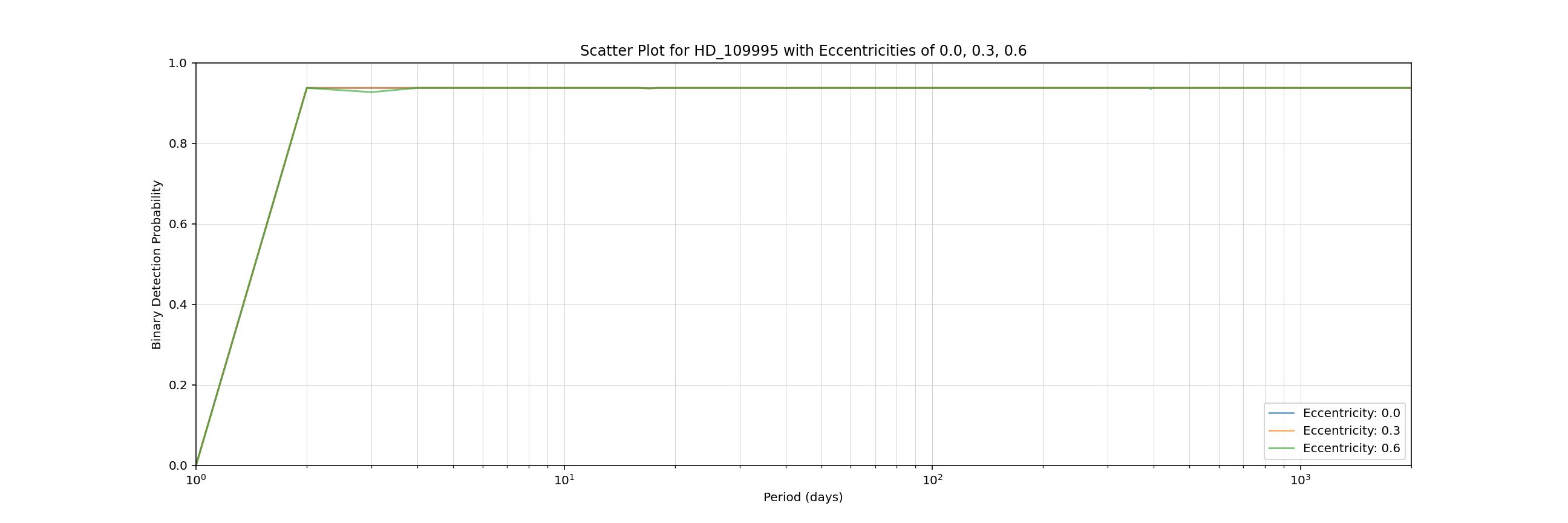}
  \includegraphics[width=\hsize]{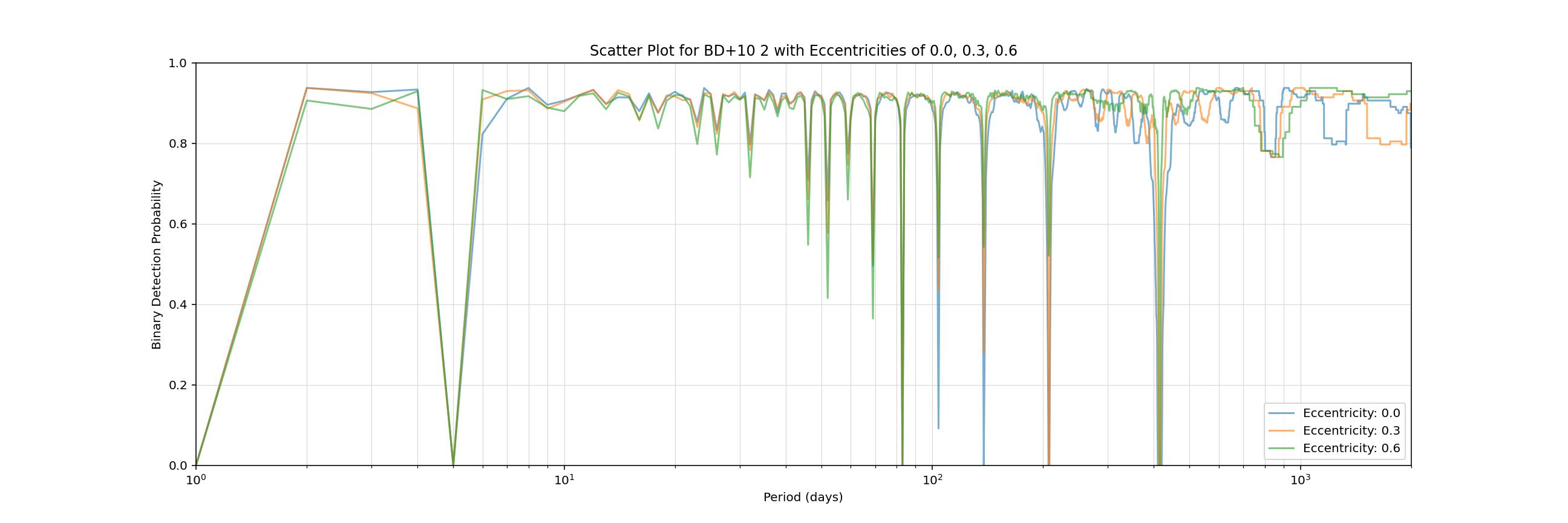}
  \includegraphics[width=\hsize]{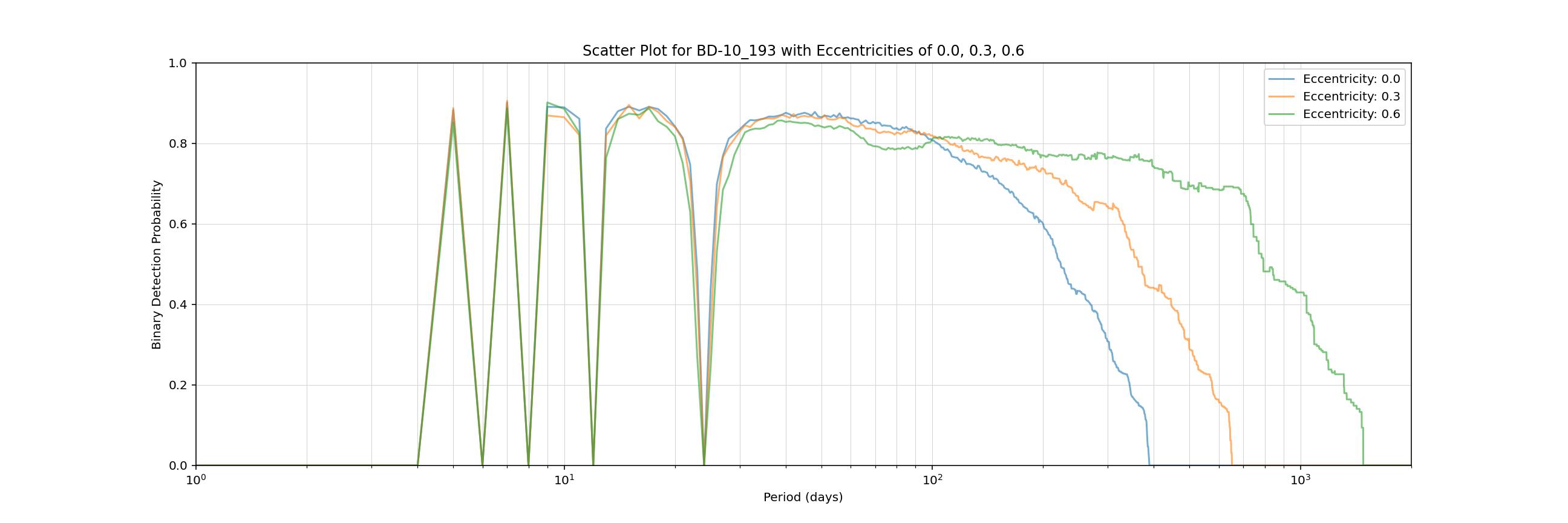}
 \caption{Binary detection probability plotted against orbital period 
 for: Top frame: BHB HD 109995 - a target with 4 spectra giving a radial velocity $v_{rad} = -129.82 \pm 0.36 $ \kms; centre frame: BHB BD+10 2 - a target with 2 spectra giving a radial velocity $v_{rad} = -267.30 \pm 1.52 $ \kms; bottom frame: MS BD-10 193 - a target with 2 spectra giving a radial velocity $v_{rad} = -102.20 \pm 7.16 $ \kms.}
  \label{binary_detection}
  \end{figure*}

In order to be able to compare results from this heterogeneous data set we calculated the weighted mean and standard error of the radial velocities for each target star. We then used these values, together with the actual time intervals between observations to calculate the probability that binarity would be detected. We found that orbital eccentricity had little effect on binary detectability rates (see Figure~\ref{binary_detection}). Thus, to assess the likelihood that a star in a binary system could evade detection due to radial velocity (RV) stability, we simulated the apparent RV variations expected from orbital motion in binaries with circular orbits as we would expect in post-RLOF binary systems. The method was tailored for systems in which the observed star dominates the flux and its companion remains undetectable as we expected for this sample and was corroborated using SED modelling.

For a binary with a primary of mass $M_1$ and an unseen secondary of mass $M_2$, in a circular orbit with period $P$, the semi-amplitude of the primary's RV is given by:

\begin{equation}
K = \left( \frac{2\pi G}{P} \right)^{1/3} \frac{M_2 \sin i}{(M_1 + M_2)^{2/3}},
\end{equation}

where $i$ is the inclination of the orbital plane and $G$ is the gravitational constant. We assumed an isotropic distribution of orbital orientations, so $\cos i$ is drawn uniformly from $[0,1]$. The orbital phase is also randomly sampled from $[0, 2\pi]$.

For each star, we generate a synthetic population of binary systems by sampling $10^5$ combinations of $P$, $i$, and phase. Orbital periods are drawn logarithmically from a uniform distribution between 1 and 5000 days. For each simulated binary, we compute the RV of the primary star at a sequence of observation times $\{t_j\}$ and evaluate the RV range $\Delta v = \max(v_j) - \min(v_j)$.

A system is considered undetected if $\Delta v < 3\sigma_{\bar{\mathrm{RV}}}$, where $\sigma_{\bar{\mathrm{RV}}}$ is the error on the weighted mean of the radial velocity $v_{\rm rad}$ (as listed in Table A1). The probability of non-detection is estimated as the fraction of simulated binaries that met this criterion.

We repeated this analysis for each star using its specific velocity resolution and observation schedule. In this way we were able to confirm that there were no intervals over the period range from 1 to 5000 days where we were unlikely to detect binarity.

BHBs in the Galactic Halo are old stars and we can assume that the majority of possible binary partners have a similar age. This limits the number of possible binary partners to being primarily main-sequence stars of a lower mass than the BHB's progenitor or white dwarfs. All other types of old star are relatively short-lived and thus should not represent a significant population. As the C24 catalogue of BHB candidates is looking at the inner Galactic Halo where star formation has ceased we have, for the purposes of selection effect determination, considered that the binary partner for a BHB will probably have a mass of around $0.8\, M_{\odot}$. White dwarfs have a mass range of $0.17\, M_{\rm \odot}\,<\,M_{\rm WD}\, < 1.33\,M_{\rm \odot}$ with the majority in the range of $0.5\, M_{\rm \odot}\,<\,M_{\rm WD}\, < 0.7\,M_{\rm \odot}$ \citep{kepler07, kepler21, obrien24}.

The lowest mass companion that we consider has $0.3\,M_{\odot}$. Since the mass transfer to such low-mass companions would likely be unstable, we expect them to have undergone a common envelope (CE) and spiral-in phase with the BHB progenitor resulting in very close post-CE BHB binaries. Since they would be very compact, they would show high RV variations. Despite their lower masses, they would therefore likely be easier to detect than more massive companions in wider orbits. Sensitivities were therefore modelled using $0.8\,M_{\rm \odot}$, $0.5\,M_{\rm \odot}$, and $0.3\,M_{\rm \odot}$ binary partners.

In this way we calculated a binary system detection probability for binary periods from 1 to 5000 days. The maximum period of a putative binary where a BHB has been stripped by RLOF is $\sim2000\,{\rm d}$ \citep{bobrick24}. The probed period interval therefore covers all types of post-interaction systems. The actual time intervals between observations for each BHB candidate observed in Ond\v{r}ejov was used to determine how likely it was that we could observe variations in apparent radial velocity arising from binary motion using the calculated radial velocities and their actual standard errors. This was done to ensure that there were no ranges of periods where we are unlikely to detect binarity using our dataset.

Given a sample of $N$ stars with no detected binary systems, we calculated upper limits on the intrinsic binary fraction (the fraction of stars in the population that are binaries) while accounting for the detection sensitivity of each observation. For each star $i$ the probability of \textit{failing} to detect a binary (if present) is denoted as $q_i = 1 - p_i$ where $p_i$ is the detection probability for that star (provided as observational constraints).

The likelihood of observing zero binaries across the sample, given a true binary fraction $f$ is the product of probabilities for each star to evade detection:
\begin{equation}
\mathcal{L}(f) = \prod_{i=1}^{N} \left( 1 - f \cdot p_i \right).
\label{eq:likelihood}
\end{equation}
Here, $1 - f$ is the probability that a star is single, and $f \cdot q_i = f \cdot (1 - p_i)$ is the probability that it is a binary but undetected.

We computed the maximum plausible binary fraction $f$ at $1\sigma$ (68.27\% confidence), $2\sigma$ (95.45\%), and $3\sigma$ (99.73\%) levels by solving:
\begin{equation}
\mathcal{L}(f) = \alpha,
\label{eq:constraint}
\end{equation}
where $\alpha$ = 1 - confidence level (e.g. $\alpha = 0.0455$ for $2\sigma$). The solution was obtained numerically via root-finding on the log-likelihood, $\mathcal{L}(f) = \sum_{i=1}^{N} ln(1 - f \cdot p_i)$ ensuring numerical stability. Upper limits in $f$ in Equation~\ref{eq:likelihood} reflected the highest rate of binarity consistent with non-detections.

Our acquired spectra, spectral analysis and synthetic SEDs have found  61 confirmed BHBs with radial velocities that are stable over time to within the errors on the RV measurements indicating that none of these objects are in binary systems. Using the above equations we calculate upper limits of BHB binarity for $0.6\, M_{\rm \odot}$ BHBs with $0.8\,M_{\rm \odot}$, $0.5\,M_{\rm \odot}$, and $0.3\,M_{\rm \odot}$ binary partners over the period ranges of 1-1000 days and 1-5000 days. Our results are shown in Table~\ref{tab:bhb_binarity} below.

\begin{table*}
    \caption{Maximum rates of binarity (to a $3\sigma$ confidence level) assuming binary partner masses of $0.8 M_{\odot}$, $0.5 M_{\odot}$ and $0.3 M_{\odot}$ and a maximum system period of 1000 days or 5000 days.}
    \centering
    \label{tab:bhb_binarity}
    \begin{tabular}{lcccccc}
    \toprule
    \toprule
    \noalign{\smallskip}
    Confidence & $P$<1000 days & $P$<1000 days & $P$<1000 days & $P$<5000 days & $P$<5000 days & $P$<5000 days \\
    level & $M_2=0.8 M_{\odot}$ & $M_2=0.5 M_{\odot}$ & $M_2=0.3 M_{\odot}$ & $M_2=0.8 M_{\odot}$ & $M_2=0.5 M_{\odot}$ & $M_2=0.3 M_{\odot}$ \\
    \noalign{\smallskip}
    \midrule
    \noalign{\smallskip}
    $1\sigma $ & 1.92\% & 2.12\% & 2.49\% & 2.20\% & 2.47\% & 2.96\% \\
    $2\sigma $ & 5.10\% & 5.63\% & 6.60\% & 5.84\% & 6.57\% & 7.84\% \\
    $3\sigma $ & 9.58\% & 10.58\% & 12.37\% & 10.98\% & 12.32\% & 14.70\% \\
    \bottomrule
    \noalign{\smallskip}
\end{tabular}
\end{table*}

We find a binarity rate of <2.2\% (for $P<5000\,{\rm d}$ and $M_2 = 0.8\,M_{\rm \odot}$) in BHBs with spectra acquired in Ond\v{r}ejov and confirmed using synthetic SEDs. This must, however, be qualified under consideration of the assumptions we have made where we have only sought binarity in BHBs with a period $P \, {\lesssim} \, 5000\,\mathrm{d}$ and we have considered binary partners with $0.3\, M_{\rm \odot} - 0.8\,M_{\rm \odot}$. As we reduce the mass of the binary partner or as we increase the period of the possible system we can be less sure of this very low rate of binarity. Our calculated rates of binarity do, however, still remain very low. 

The additional UVES and FEROS spectra had observation dates between 2001 and 2018 thus extending the timeline of our radial velocity data points from 3 years considerably to between 7 and 22 years but only for a very low number of stars. The observation dates, radial velocities (with barycentric corrections applied) and radial velocity errors for all BHB candidate stars observed in Ondrejov are listed in Appendix A.
\section{Detection of binarity using light curve variations}
\label{sect:photometric}

We used the \textit{Gaia} DR3 and ZTF datasets to try to find photometric variations in the C24 BHB candidates. Eclipses are more likely to be detected in systems where the binary partners are closer. Consequently, only a limited fraction of such systems will be detected in this way. By calculating the range of angles over which an observer will perceive eclipses (see equation~\ref{eq:eclipsing_angle}) we calculated the eclipse detection probability. 
\begin{equation}
\label{eq:eclipsing_angle}
    \Delta \phi = 2 \times \arctan\left(\frac{D_{\text{total}}}{2a}\right),
\end{equation}

where \( D_{\text{total}} = D_1 + D_2 \) is the sum of the diameters of the binary partners and \( a \) is the orbital separation between the binary partners.

Using the assumptions made in Section 2.4 ($M_{\rm BHB}  \, {\simeq}  \, 0.6 \rm M_{\odot}$ and $M_{\rm MS}  \, {\simeq}  \, 0.8 \rm M_{\odot}$) together with radii of BHBs ranging from $1\, R_{\rm \odot} < R_{\rm BHB} < 4\,R_{\rm \odot}$ and $R_{\rm MS}  \, {\simeq}  \, 0.74\,R_{\rm \odot}$ as given in BaSTI isochrones \citep{pietrinferni04} we see that the probability of observing an eclipse drops to around 1\% when the binary separation reaches $1\,{\rm au}$. Such a binary system would have an orbital period of about 300 days. Photometric detection of binary systems from eclipses is strongly biased towards close binary systems.

Binarity can also be detected by finding periodic photometric variations due to ellipsoidal distortion or colour asymmetry. The angle between the observer and the orbital plane of the binary system is less restrictive than for observing eclipses but there is an even stronger bias toward close binary partners. Such effects can only be observed for periods up to a few days. 

\subsection{\textit{Gaia} DR3 time-variable data}

\textit{Gaia} DR3 contains a time-variability analysis that identifies 10.5 million \textit{Gaia} DR3 objects as variable. This analysis is described in \citet{eyer17} and the results of the analysis are given in \citet{eyer23}. The \textit{Gaia} DR3 variability is split into 11 categories. $G$-band light curve modelling for eclipsing binary light curves combines Gaussian functions and sine functions to model eclipsing and ellipsoidal deformation respectively \citep{mowlavi23}. {\em Gaia} has an exposure time of 4.4 seconds per CCD and 34 months of data have been processed for {\em Gaia} DR3 time-variability analysis. These figures drive the limits of the period range that can be found using these data.

We crossmatched the C24 BHB candidates with the \textit{Gaia} DR3 variability catalogue and found 2020 objects had released light curves of which 113 stars showed variability typical for eclipsing binaries. Only \textit{Gaia} DR3 targets that showed significant variability in their photometric data have their light curves released.

However, CMD plots reveal systematic shifts between the eclipsing binary candidates and the BHB candidate population. Based on colours and absolute magnitudes of the spectroscopically confirmed BHBs (see Sect. 2) we defined the core BHB region (to $1\sigma$ and $2\sigma$ uncertainty) using a kernel density estimate. For comparison we also defined the core CMD region (to $1\sigma$ and $2\sigma$ uncertainty) of the Gaia DR3 eclipsing binaries (EB) (see Table~\ref{tab:kde_population}).

 \begin{table}
    \caption{Presence of C24 and Gaia DR3 EB stars in KDE regions}
    \centering
    \label{tab:kde_population}
    \begin{tabular}{lcccc}
    \toprule
    \toprule
    \noalign{\smallskip}
    Population & BHB & BHB & EB & EB  \\
    KDE Uncertainty & $1\sigma$ & $2\sigma$ & $1\sigma$ & $2\sigma$ \\
    \noalign{\smallskip}
    \midrule
    \noalign{\smallskip}
    C24 Candidates & 4,127 & 11,346 & 5,596 & 12,012 \\
    \textit{Gaia} DR3 EB & 1 & 26 & 66 & 101 \\
    \bottomrule
    \noalign{\smallskip}
\end{tabular}
\end{table}

\begin{figure*}[t]
  \centering
  \includegraphics[width=\hsize]{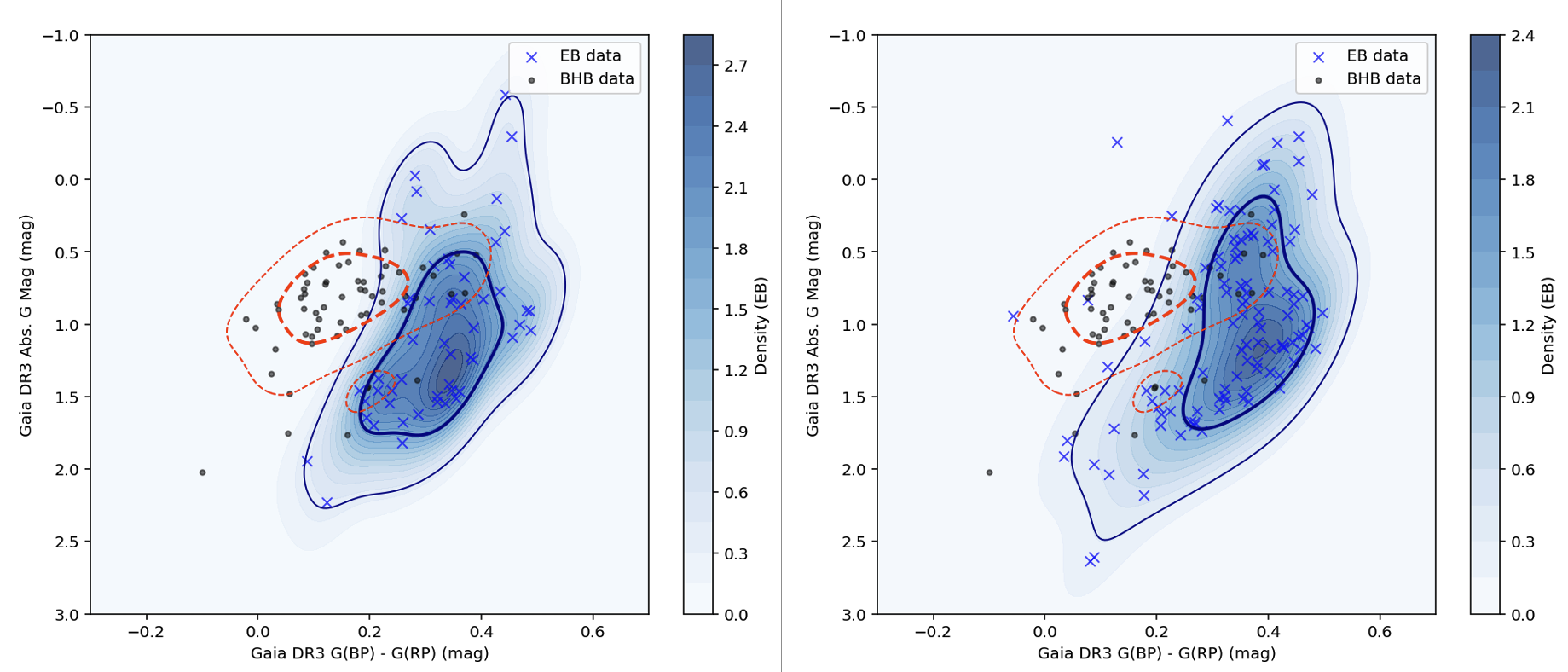}
 \caption{Gaussian kernel density estimate (KDE) CMD plots for the ZTF eclipsing binary objects (left panel) and \textit{Gaia} DR3 eclipsing binary objects (right panel) found in the C24 catalogue (blue crosses on both panels) with the 1$\sigma$ and 2$\sigma$ contours superimposed. Spectrally confirmed BHB candidates (gray dots on both panels) with the 1$\sigma$ and 2$\sigma$ contours (red) superimposed (both panels).}
  \label{kde}
  \end{figure*}

The \textit{Gaia} DR3 selection criteria for publishing light curve data included cutoffs for very bright stars due to the possibility of saturation and very faint stars due to the poor signal-to-noise ratio. The 113 C24 BHB candidates that showed \textit{Gaia} DR3 variability typical of eclipsing binaries lie within the apparent $G$ magnitude range of 12.0 - 16.8 mag and the parallax range of 0.11 - 0.70 mas giving a distance range of 1.4\,kpc to 9.0\,kpc. 21,546 (96.5\%) of the C24 BHB candidates also lie within these ranges.

\citet{mowlavi23} states that the  \textit{Gaia} DR3 eclipsing binaries are 25\%-50\% complete depending on the region on the sky and complete out to 3-5\,kpc for brighter targets. Beyond $\sim3-5$\,kpc, incompleteness becomes significant due to fainter systems being missed, crowding in dense regions, and limited photometric sensitivity for low-amplitude variations. Although not quantified here none of these limiting criteria are applicable to the C24 catalogue. Hence conservatively considering the \textit{Gaia} DR3 eclipsing binary catalogue to be complete out to 5\,kpc where 13,118 (42\%) of the C24 BHB candidates and 59 of the 113 cross-matched \textit{Gaia} DR3 eclipsing binaries are found indicates that we are not missing a significant number due to selection effects. 

We plot the 61 confirmed BHBs together with the 113 \textit{Gaia} DR3 eclipsing binaries found in the C24 catalogue on a CMD and generated Gaussian kernel density estimates for the 2 datasets (see Figure~\ref{kde} right panel). This showed a clear distinction between the confirmed BHBs and the \textit{Gaia} DR3 eclipsing binaries. The small number of EBs in the core BHB region in contrast to the systematically shifted and much larger numbers of EBs in CMD regions, where MS-A/B stars and blue stragglers become more abundant, leads us to the conclusion that most of the \textit{Gaia} DR3 eclipsing binaries that are found in the C24 catalogue are likely to be contaminants (see Table A1).

\subsection{Identifying binarity with ZTF data}

ZTF targets result from scanning the northern sky in the g and r filters for all variable phenomena greater than ${\lesssim} 20.5$\,mag on timescales of minutes to years \citep{graham19}. The ZTF field of view corresponds to most of the region with declination < $-$28°. Thus, 12,114 (54\%) of the 22,336 C24 BHB candidates 
have been scanned by ZTF. Of these objects, 52 were found to be periodic variables in the ZTF dataset by \citet{chen20},  who used ZTF DR2 as their input dataset. 

As was seen in the \textit{Gaia} DR3 eclipsing variables, when plotting the ZTF variables on CMD plots (see Figure~\ref{kde} left panel) we can see that these objects are concentrated away from the central region where most of the spectroscopically confirmed C24 BHB candidates cluster giving a further indication that the C24 BHB candidates that show variable photometric behaviour are likely to be contaminants in the C24 BHB catalogue.


\section{Astrometric detection of binarity}
\label{sect:astrometric}

Over 96\% of C24 have a distance as calculated from parallax of ${\gtrsim} 2\,{\rm kpc}$. \textit{Gaia} DR3 is expected to be able to resolve stars down to a separation of 0.1 arcsec. Thus, we calculate that binary systems with a separation of less than $200\,{\rm au}$ cannot be resolved for the range of parallaxes in C24. In order to be able to detect binarity in systems closer than this limit, we can only detect binary motion by observing the BHB candidate star and making the assumptions from Section 2.1 for the binary partner.

High rates of binarity in BHBs were predicted by \citet{belokurov20} section 3.4 based on the \verb!ruwe! factor calculated from the astrometry of {\em Gaia} DR2. \citet{belokurov20} showed that the \verb!ruwe! factor is sensitive to astrometric "photocentre wobble" that is typical for unresolved binaries.

\subsection{\textit{Gaia} DR3 renormalised unit weight error}

\begin{figure}
  \centering
  \includegraphics[width=\hsize]{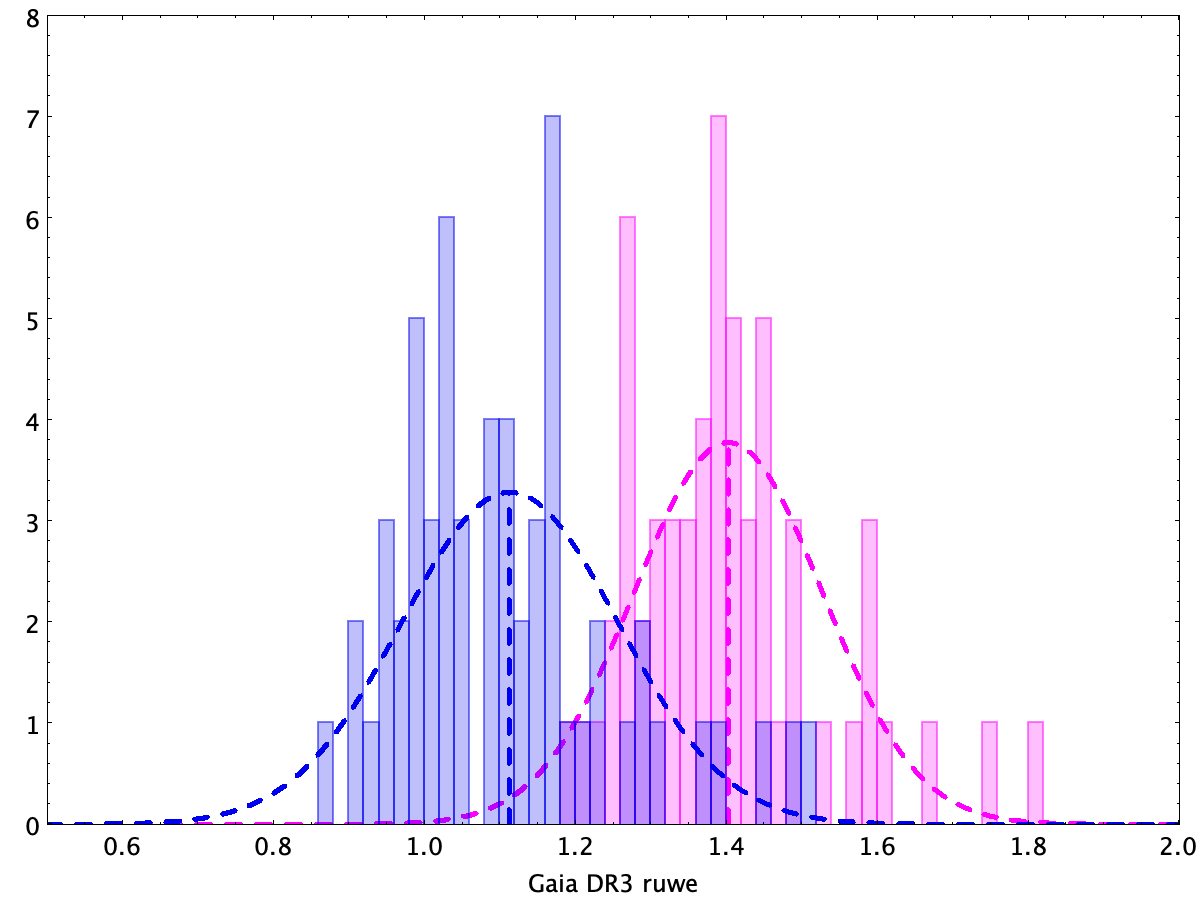}
 \caption{The best-fit Gaussian distributions for $ruwe$ of 61 spectroscopically confirmed BHBs (blue) and the mean $ruwe$ for 500 observationally similar targets (magenta) for each confirmed BHB.}
  \label{ruwe}
  \end{figure}

We performed a similar exercise to \citet{belokurov20} using the 61 BHBs that were confirmed using Ond\v{r}ejov spectra and synthetic SEDs. We compared the \textit{Gaia} DR3 \verb!ruwe! factor (as opposed to \textit{Gaia} DR2 used by \citet{belokurov20}) for the confirmed BHBs to 500 (as opposed to 30 used by \citet{belokurov20}) observationally similar \textit{Gaia} DR3 objects. We did this by finding the 500 nearest targets in terms of colour (\verb!bp_rp!), apparent $G$ magnitude (\verb!phot_g_mean_mag!), $G$ flux error (\verb!phot_g_mean_flux_over_error!) and the number of observations (\verb!phot_g_n_obs!) while ensuring that the same quality criteria for parallax error, the astrometric solution, and the colour excess were also applied. \citet{belokurov20} considered colour (\verb!bp_rp!), and apparent $G$ magnitude (\verb!phot_g_mean_mag!) in identifying observationally similar targets. The mean \verb!ruwe! of these 500 observationally similar targets were compared to the \verb!ruwe! factor of the 61 confirmed BHBs. We found that the confirmed BHBs had \verb!ruwe! values that were consistently lower than observationally similar targets (see Figure~\ref{ruwe}) in contrast to the findings in \citet{belokurov20}.

We found a similar distribution of \verb!ruwe! for our 61 spectroscopically confirmed BHBs as \citet{belokurov20} found for their 54 BHB candidates whereas our observationally similar targets showed higher \verb!ruwe! values than were found by \citet{belokurov20}. In conclusion, our study indicates a smaller astrometric binary fraction for the BHBs compared the control sample consistent with our results using the other methods.

\subsection{\textit{Gaia} DR3 common proper-motion partners}

\begin{figure}
  \centering
  \includegraphics[width=\hsize]{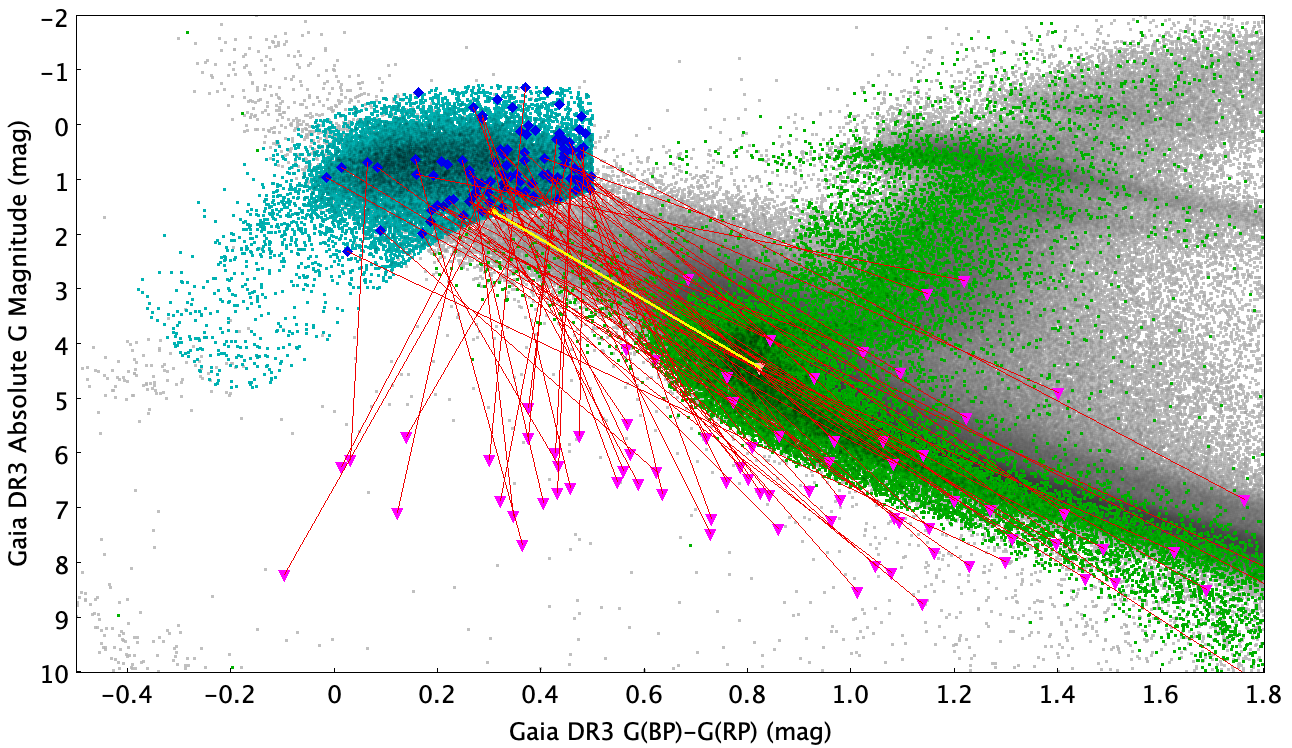}
  \includegraphics[width=\hsize]{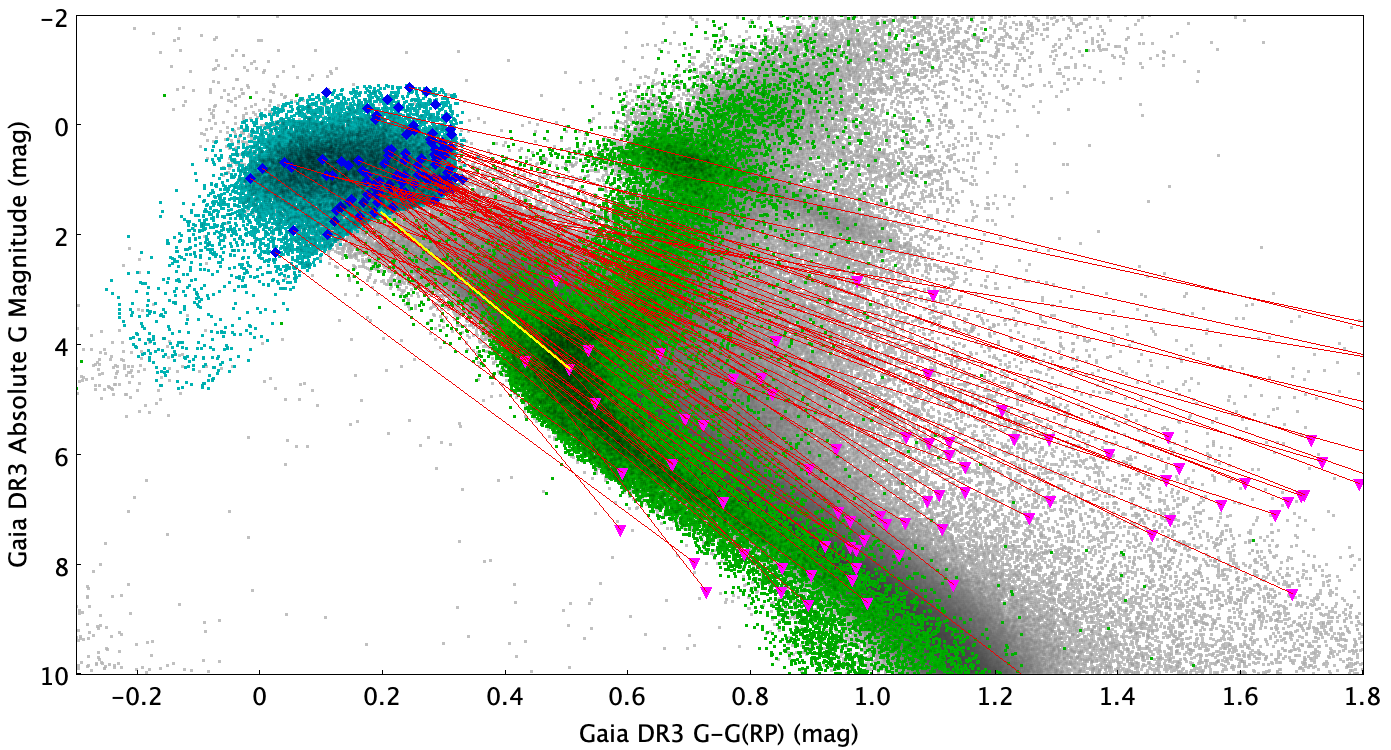}
 \caption{Upper frame: \textit{Gaia} DR3 CMD with $G_{BP}- G_{RP}$ colour; lower frame: \textit{Gaia} DR3 CMD with $G - G_{RP}$ colour;  for 2,000,000 randomly selected objects with parallax errors less than 10\% (grey dots), and the subset of halo candidates with tangential velocities greater than 145 \kms\ (green dots). C24 BHB candidates (cyan dots). BHB candidates with CPM partner candidates (blue triangles), and CPM partner candidates (magenta diamonds). The only BHB candidate - CPM partner candidate pair where both pass the parallax error < 20\% criterion are joined with a bold yellow line in both the upper and lower frames.}
  \label{cmd_cpm}
  \end{figure}

A further method of detecting binaries using astrometric data is searching for common proper motion (CPM) candidates as was done in \citet{pelisoli20}. We used the ADQL script given in Appendix B which was based upon the script used by \citet{pelisoli20} to find CPM partners to hot subdwarfs. We found only 118 of the 22,336 BHB candidates from C24 had CPM partner candidates. It should be noted, however, that for a $200\,{\rm au}$ separation we can only expect to resolve the partners up to 2\,kpc (assuming 0.1 arcsec \textit{Gaia} DR3 resolution) corresponding to only 821 C24 candidates. This increases to 10\,kpc for a separation of 1,000 au meaning that CPM partners for the entire C24 could be observed at this separation. Only one CPM partner candidate conformed to the same parallax quality criteria (\verb!parallax! > 0, \verb!parallax_over_error! > 5). All others have less reliable parallaxes. When comparing the 118 CPM pairs on two \textit{Gaia} DR3 CMDs, one using the G(BP)-G(RP) colour and the other using the G-G(RP) colour we can see that the majority of objects have unreliable colours and plot on different regions of their respective CMDs. The one BHB candidate whose CPM partner candidate had satisfactory parallax plots in the same CMD region in both diagrams (see Figure~\ref{cmd_cpm}).

This very low number of candidates was potentially caused by one of the C24 BHB candidate selection criteria that required that BHB candidates with apparent neighbours within 5 arcsec must be the source of >70\% of the G-band flux coming from the 5 arcsec radius window around the candidate itself. BHBs are, however, very bright objects with the C24 candidates having a mean absolute magnitude of 0.57 mag with a standard deviation of 0.62 mag. Thus, this selection effect will only apply to the similarly bright common proper motion partners. As we can exclude higher mass main-sequence stars due the age limitations in the Galactic halo this leaves us with relatively short lived red giants. White dwarfs and lower mass main-sequence CPM partners will not be lost due to this selection effect.

Such CPM partner candidates with lower luminosities do, however, have less well determined astrometric and photometric data as their apparent proximity to the bright BHBs is detrimental in the determination of their position and apparent magnitude.

The one BHB candidate (\textit{Gaia} DR3 3537410759489233792) with a CPM candidate that satisfied the parallax QC criteria was further analysed by generating a synthetic spectral energy distribution (SED) as described in Section 2.1 above. This found a best-fit SED for a star with $T_\mathrm{eff}=8280\pm40\rm\,K$, $\log~g=4.01^{+0.12}_{-0.14}$, $R=2.25^{+0.09}_{-0.08}\,R_{\rm \odot}$, $M=1.9\pm0.6\,M_{\rm \odot}$, and $L=21.3^{+1.7}_{-1.6}\,L_{\rm \odot}$ indicating it is a main-sequence star. It is highly unlikely that a $M=1.9\pm0.6\,M_{\rm \odot}$ MS star can be the binary partner of a halo BHB hence this is either a chance alignment or the BHB candidate is actually a contaminant.

\section{Summary and conclusions}
\label{sect:summary}

In this work we carried out the first systematic, multi-technique search for binarity among blue horizontal-branch (BHB) stars in the inner Galactic halo, combining spectroscopy, astrometry, and photometry. Starting from the C24 catalogue of 22,336 BHB candidates from \textit{Gaia} DR3, we obtained 263 spectra of 89 targets with the Ondřejov Echelle Spectrograph and supplemented these with archival UVES and FEROS data, yielding a total of 319 high-resolution spectra for radial velocity monitoring. Our spectroscopic analysis probes binarity up to orbital periods of 5000 days. Our observations are supported by crossmatching the C24 catalogue of BHB candidates with \textit{Gaia} DR3 eclipsing binaries, and ZTF datasets where binary light curve variability probes the short period of the order of a few days.

We further examined Gaia astrometry for excess noise where we would expect a higher RUWE factor for periods of the order of 1000 days. We also probe a significant part of the longer period range, where non-interacting binaries are usually found by searching for common proper-motion companions.

When considering the datasets we have used in conjunction with Figure 1 from \citet{moe17} where binarity detection efficiency using different detection methods are plotted against the binary orbital period we can see that we are operating in the region where stars in binary systems that interact through Roche-lobe overflow (RLOF) should be detected. As shown in Section 11 of \citet{moe17} binary systems with a solar-type member will interact through RLOF when the orbital period is less than about 10 years.

Synthetic SED modelling was used to identify hitherto undetected contaminants and tailored simulations were used to account for detection biases and selection effects. Across all methods we find no significant evidence for binarity: both close systems detectable via astrometric excess noise or radial-velocity variations and wide systems revealed through common proper-motion pairing are absent. The C24 targets found to show binary light curve variability are more likely to be contaminants as they plot coherently away from the main BHB cluster on the CMD. Despite the much larger sizes of these datasets, we find no indications of a population of binaries exceeding the upper limit derived from our smaller spectroscopic sample. We therefore place a stringent $1\sigma$ upper limit of $<2.2\%$ on the binary fraction (P < 5,000 days; partner mass $0.8\,M_{\rm \odot}$) of halo BHB stars. Our findings align well with those from \citet{green14} who looked at radial velocity variations in 10 field BHB stars as well as the findings from \citet{monibidin11} who looked at 64 BHB and EHB targets with $T_{\rm eff} > 12,000\,{\rm K}$ in globular clusters (GCs).\footnote{A significantly higher binary fraction was reported recently by \citet{guo25} based on an RV study using medium-resolution spectra of a spectroscopically selected sample of field BHB stars \citep{ju24}. The high discrepancy to our and other published results might be caused by differences in filtering out contaminants. Applying the synthetic SED method outlined in Section 2.3 here could resolve this issue.}

Given these low rates of binarity among halo BHB stars, it is essential to place our results in the broader context of other horizontal-branch populations and their common progenitors. Comparisons with the well-studied binary fractions of main-sequence, red-giant, and extreme horizontal-branch stars, as well as related populations such as RR Lyrae and red clump stars, provide crucial insights into whether the lack of companions we observe is unique to BHBs or reflects a more general trend across different evolutionary channels.

Low-mass, metal-poor main-sequence stars are progenitors of halo BHB stars. Their binary properties provide the natural baseline for comparison. Surveys of main-sequence populations have shown that the close-binary fraction is strongly anti-correlated with metallicity \citep{moe19}: while solar-metallicity MS stars display fractions of $\sim20-30\%$, stars at $[{\rm Fe/H}] \lesssim -1$ exhibit markedly higher rates, up to $40-50\%$ for periods shorter than $10^4$ days. These elevated binary fractions at low metallicity are especially relevant in the context of BHB progenitors, since they imply that close companions should have been common in the early halo population.

As main-sequence stars evolve onto the red-giant branch, binary interactions become increasingly important, since close systems can undergo Roche-lobe overflow or common-envelope evolution. The observed binary fractions of red giants remain high, typically $\sim$30–50\% depending on metallicity \citep{moe19}, but the distribution is skewed toward wider systems because the closest binaries have already interacted and in many cases merged or produced stripped remnants. Since every BHB star must have passed through a red-giant phase, their binary incidence would be expected to resemble that of halo giants - reduced compared to their main-sequence progenitors in the very closest regimes, but still substantial overall. The stark contrast with the very low binary fraction we measure for BHBs therefore requires closer examination.

\begin{table*}
  \centering
  \caption{Published binary fractions for progenitors of blue horizontal-branch stars.}
  \label{tab:binary_rates}
  \begin{tabular*}{\textwidth}{@{\extracolsep{\fill}}lcccc@{}}
    \toprule
    \toprule
    Population & Observed & Selection & Metallicity [Fe/H] & Reference \\
      & Binary Fraction & Effect Corrected$^{*}$ &   &   \\
    \midrule
    BHB (this work) & $0\%$ & $<2.2\%$ & $-1.7 \pm 0.3$ & -- \\
    RGB & ${\sim} 18\pm6\%$ & ${\sim} 57\pm22\%^{*}$ & $< -3.0\pm0.7$ & \citet{hansen15} \\
    RGB & ${\sim} 18\pm4\%$ & ${\sim} 34\pm12\%^{*}$ & $< -2.0\pm0.7$ & \citet{carney03} \\
    MS & ${\sim} 15-20\%$ & $54\pm12\%^{*}$ & -2.7 & \citet{latham02,goldberg02} \\
    MS & $14\pm2\%$ & $20\pm3\%^{*}$ & 0 & \citet{latham02,goldberg02} \\
    \bottomrule
  \end{tabular*}
  \vspace{2mm}
  \small *selection effect corrections to P=10,000 days; a < 10 AU from \citet{moe19}
\end{table*}

Among adjacent horizontal-branch populations, markedly different binary fractions are observed. RR Lyrae stars in the halo show only a few confirmed binaries \citep{bobrick24} implying fractions of at most a few percent. This is consistent with halo EHB stars, which show neither post-CE nor any confirmed post-RLOF systems \citep{latour18,geier24} thus small close binary fractions seem to be a general feature of HB stars in the halo. This is, however, in stark contrast to the high observed close binary fractions of EHB stars in the Galactic thin disc \citep[e.g.][]{geier22} and theoretical predictions according to which binary interactions are needed to form RR\,Lyrae and RHB stars in such young stellar populations \cite{matteuzzi23,bobrick24}.

It is possible that BHBs can only form when no binary interaction has taken place. It is also possible that low mass MS stars can only evolve to BHBs with a binary partner that is lost from the system in some way (supernova, dynamic ejection, merger or similar). Metal-poor binaries are more likely to merge during the RGB phase due to shorter orbital decay timescales \citep{vanrensbergen20}. This could explain the scarcity of detected BHB binaries in the inner Galactic Halo as well as the lack of wider, non-interacting, binaries among the BHBs in the inner Galactic Halo.

Either way, our research indicates strongly that binarity is a critical factor in the stellar evolution of BHB progenitors as they evolve from the MS through the RGB and onto the BHB. Understanding which of these possible scenarios is the most likely requires further work to develop a catalogue of younger, higher metallicity BHB candidates which can then be subject to a similar analysis.

\begin{acknowledgements}

We thank the following workshop attendees for their diligence in acquiring the spectra used in this project. 2021 workshop: Ayesha Arshad Arain, Saksham Arora, Rahda Anil Gharapurkar, Siddarth Khalate, Chinmay Mahajan, Henrik Rose, Amrit Sedain, Tahereh Ramezani, Xia Caiyun, Prapti Mondal and Kate\v{r}ina Pivo\v{n}kov\'a. 2022 workshop: Shubham Mamgain, Ajinkya Kakade, Prem Kumar, Saqib Sumra Muhammad, Sreepriya Vijayasree, Sahil Jhawar, Aleeda Charly, Cinta Vidante, Pouria Samieadel, Samaneh Zahmatkeshfilabi, Ravi Shankar Chaurasia, Vivek Reddy Pininti, Sabzali Vajihe, Michael V\'avra, Vojt\v{e}ch Partík, Semír Aldabagh, and Alexander Dimoff. 2023 workshop: Ramona Valkov\'a, Hemish Delvadiya, Artem Gorodilov, Aman Mohan, Ramon Jaeger, Henry Willems, Kadir Tolkay Uludag, Gitanjali Gitanjali, Shefali Negi, Arpitha Paramel Velayudhan, Krist\'yna Janou\v{s}kov\'a, and Henry Willems.

We thank Andreas Irrgang (FAU Erlangen-N\"urnberg) for amending the ATLAS12 code and developing the SED fitting tool.

We thank Marilyn Latour for her supporting material and comments.

We thank Vasily Belokurov for his comments.

M.D. was supported by the Deutsches Zentrum für Luft- und Raumfahrt (DLR) through grant 50-OR-2304. 

I.P. acknowledges funding by the UK’s Science and Technology Facilities Council (STFC), grant ST/T000406/1, and from a Warwick Astrophysics prize post-doctoral fellowship made possible thanks to a generous philanthropic donation. I. P. was supported by the Deutsche Forschungsgemeinschaft (DFG) through grant GE2506/12-1.

A.B., M.P. and F.M. were supported by the DFG through grant GE2506/18-1. H.D. was supported by the DFG through grants GE2506/12-1, GE2506/17-1 and GE2506/9-2. 

B.K. and J.K. acknowledge the support from the Mobility Plus project between the Czech Academy of Sciences and
Deutscher Akademischer Austauschdienst  (DAAD-24-02). The Astronomical Institute of the Czech Academy of Sciences (ASU) in Ond\v rejov is supported by the project RVO: 67985815. We thank the Stellar Physics Department of ASU and their technical staff for the support during the observations.

This research made use of TOPCAT, an interactive graphical viewer and editor for tabular data Taylor (\cite{taylor05}). This research made use of the SIMBAD database, operated at CDS, Strasbourg, France; the VizieR catalogue access tool, CDS, Strasbourg, France. Some of the data presented in this paper were obtained from the Mikulski Archive for Space Telescopes (MAST). STScI is operated by the Association of Universities for Research in Astronomy, Inc., under NASA contract NAS5-26555. Support for MAST for non-HST data is provided by the NASA Office of Space Science via grant NNX13AC07G and by other grants and contracts. This research has made use of the services of the ESO Science Archive Facility.

This work has made use of data from the European Space Agency (ESA) mission {\it Gaia} (https://www.cosmos.esa.int/gaia), processed by the {\it Gaia} Data Processing and Analysis Consortium (DPAC, https://www.cosmos.esa.int/web/gaia/dpac/consortium). Funding for the DPAC has been provided by national institutions, in particular the institutions participating in the {\it Gaia} Multilateral Agreement.

This work has made use of BaSTI web tools.

\end{acknowledgements}

\clearpage
\begin{onecolumn}
\begin{appendix}

\section{List of targets with spectra acquired in Ond\v{r}ejov}

\begin{longtable}{lcccccccc}
\caption{\label{table:A1} Spectral analysis results}\\
\hline\hline
\noalign{\smallskip}
Simbad id & Type & OES & UVES/ & BJD & $v_\mathrm{rad}$ & $\sigma_{v_{rad}}$ & \vsini\ $\pm  1\sigma$  & $v_\mathrm{rad}$ $\pm  3\sigma$ \\
 & & spectra & FEROS &   & & & weighted mean & weighted mean  \\
  &  &  & spectra &  & (\kms) & (\kms) & (\kms) & (\kms) \\
\noalign{\smallskip}
\hline
\noalign{\smallskip}
BD-10 193 & BHB & 2 & 0 & 2460196.58347965 & -103.13 & 3.28 & $38.5\pm2.0$ & $-102.0\pm7.2$ \\
 &  &  &  & 2460219.49502417 & -100.85 & 3.38 &  & \\
BD-10 946 & BHB & 2 & 0 & 2458510.03887914 & -22.45 & 0.47 & $24.2\pm2.0$ & $-21.5\pm0.9$ \\
 &  &  &  & 2459947.37342738 & -21.17 & 0.36 &  & \\
 &  &  &  & 2460295.38661139 & -23.22 & 0.87 &  & \\
BD+00 145 & BHB & 2 & 0 & 2459804.59448787 & -249.88 & 1.39 & $34.7\pm4.0$ & $-256.6\pm2.4$ \\
 &  &  &  & 2460219.51912426 & -259.85 & 0.98 &  & \\
BD+01 548 & BHB & 2 & 3 & 2489051.87928421 & -53.64 & 0.51 & $28.7\pm5.2$ & $-53.6\pm1.5$ \\
 &  &  &  & 2457716.92197728 & -52.12 & 1.48 &  & \\
 &  &  &  & 2459093.06281629 & -52.90 & 0.62 &  & \\
 &  &  &  & 2459947.24154368 & -53.29 & 0.86 &  & \\
 &  &  &  & 2460012.27054955 & -56.88 & 1.17 &  & \\
BD+03 4247 & BHB & 2 & 1 & 2489102.11699180 & -30.97 & 0.25 & $19.6\pm9.8$ & $-31.3\pm0.7$ \\
 &  &  &  & 2460140.48847709 & -35.52 & 0.98 &  & \\
 &  &  &  & 2460197.45584954 & -38.73 & 21.82 &  & \\
BD+07 2360 & MS & 3 & 0 & 2459641.49215745 & 103.35 & 1.10 & $21.3\pm1.8$ & $103.7\pm1.5$ \\
 &  &  &  & 2459642.54320161 & 103.81 & 0.53 &  & \\
 &  &  &  & 2460068.34241241 & noisy & - &  & \\
BD+10 2 & BHB & 2 & 0 & 2459803.56862487 & -267.03 & 2.12 & $24.8\pm2.1$ & $-267.3\pm1.5$ \\
 &  &  &  & 2460218.47451707 & -267.32 & 0.55 &  & \\
BD+15 34 & BHB & 4 & 0 & 2459803.50334855 & -78.62 & 0.42 & $26.0\pm2.3$ & $-79.0\pm0.3$ \\
 &  &  &  & 2459821.50338154 & -78.13 & 2.04 &  & \\
 &  &  &  & 2460246.3924304 & -18.20 & 1.49 &  & \\
 &  &  &  & 2460295.22587751 & -79.75 & 0.53 &  & \\
BD+18 3429 & MS & 3 & 0 & 2459466.33242001 & -33.99 & 4.13 & $287.8\pm3.3$ & $-27.4\pm3.6$ \\
 &  &  &  & 2459803.31707704 & -26.64 & 1.35 &  & \\
 &  &  &  & 2459821.36435224 & -27.53 & 2.51 &  & \\
BD+25 2602 & BHB & 3 & 4 & 2475500.16890549 & -64.57 & 2.35 & $20.9\pm1.0$ & $-66.5\pm0.9$ \\
 &  &  &  & 2476520.57353441 & -66.64 & 1.19 &  & \\
 &  &  &  & 2475810.11582864 & -63.82 & 3.40 &  & \\
 &  &  &  & 2475301.20049587 & -66.17 & 0.41 &  & \\
 &  &  &  & 2459663.44245589 & -66.17 & 1.28 &  & \\
 &  &  &  & 2460021.50828488 & -66.69 & 0.56 &  & \\
 &  &  &  & 2460050.3805212 & -66.55 & 0.44 &  & \\
BD+25 4066 & BHB & 1 & 0 & 2460195.36049083 & -143.95 & 0.70 & $24.7\pm9.0$ & $-144.0\pm2.1$ \\
BD+29 2058 & MS & 3 & 0 & 2459642.575647 & 153.89 & 3.21 & $287.1\pm3.2$ & $157.8\pm6.6$ \\
 &  &  &  & 2459699.45576584 & 159.86 & 6.12 &  & \\
 &  &  &  & 2460088.33444943 & 161.55 & 3.38 &  & \\
BD+36 3268 & BHB & 3 & 0 & 2459772.51635186 & -95.91 & 0.51 & $16.8\pm1.2$ & $-96.0\pm1.2$ \\
 &  &  &  & 2459826.27974808 & -95.79 & 0.79 &  & \\
 &  &  &  & 2460198.5453546 & -96.35 & 0.58 &  & \\
BD+38 574B & BHB & 4 & 0 & 2459468.53565464 & -102.05 & 2.28 & $24.5\pm1.4$ & $-105.6\pm1.2$ \\
 &  &  &  & 2459470.46720177 & -101.60 & 0.68 &  & \\
 &  &  &  & 2459828.55239564 & -102.07 & 1.21 &  & \\
 &  &  &  & 2460295.27298417 & -101.44 & 0.43 &  & \\
BD+40 5154 & BHB & 3 & 0 & 2459782.57540171 & -260.30 & 2.25 & $25.5\pm2.3$ & $-262.8\pm2.1$ \\
 &  &  &  & 2459824.56481966 & -260.02 & noisy &  & \\
 &  &  &  & 2459947.32556986 & -263.04 & 0.71 &  & \\
BD+42 2309 & BHB & 2 & 0 & 2459663.54301007 & -142.39 & 0.93 & $38.5\pm2.0$ & $-142.3\pm1.8$ \\
 &  &  &  & 2460014.30791672 & -140.81 & 3.62 &  & \\
BD+43 2344 & BHB & 2 & 0 & 2459651.52450499 & -149.05 & 0.65 & $28.6\pm4.0$ & $-149.0\pm1.8$ \\
 &  &  &  & 2460061.36124151 & -148.49 & 1.67 &  & \\
BD+44 2523 & MS & 3 & 0 & 2459651.68623239 & noisy & - & $285.5\pm3.8$ & $-82.3\pm5.2$ \\
 &  &  &  & 2460296.63835464 & -79.88 & 1.98 &  & \\
 &  &  &  & 2460330.60576885 & -84.95 & 2.07 &  & \\
BD+83 337 & MS & 2 & 0 & 2459471.60000846 & noisy & - & $442.2\pm5.8$ & $49.2\pm7.8$ \\
 &  &  &  & 2459802.3833844 & 49.20 & 2.62 &  & \\
HD 101292 & BHB & 2 & 0 & 2459675.3865035 & -54.45 & 0.54 & $26.1\pm3.7$ & $-54.7\pm1.2$ \\
 &  &  &  & 2459692.35556379 & -55.08 & 0.71 &  & \\
HD 104607 & BHB & 2 & 0 & 2459663.39864081 & 83.97 & 0.51 & $38.4\pm1.7$ & $83.9\pm1.5$ \\
 &  &  &  & 2460061.31984068 & 83.61 & 0.92 &  & \\
HD 109995 & BHB & 4 & 4 & 2459657.54127291 & -129.68 & 0.20 & $27.3\pm0.8$ & $-129.8\pm0.3$ \\
 &  &  &  & 2460014.48502734 & -130.15 & 0.31 &  & \\
 &  &  &  & 2460020.44987417 & -129.63 & 1.11 &  & \\
 &  &  &  & 2460054.45724284 & -129.81 & 0.17 &  & \\
HD 110012 & MS & 2 & 0 & 2459663.62937486 & -63.29 & 1.36 & $165.4\pm3.9$ & $-60.5\pm3.0$ \\
 &  &  &  & 2459980.45267417 & -57.09 & 1.51 &  & \\
HD 122190 & BHB & 2 & 1 & 2475584.31481704 & 156.45 & 0.42 & $20.9\pm1.5$ & $156.4\pm1.2$ \\
 &  &  &  & 2459748.35132372 & 156.13 & 0.94 &  & \\
 &  &  &  & 2460088.42743655 & 156.35 & 1.21 &  & \\
HD 132471 & BHB & 2 & 1 & 2482809.0860738 & 118.40 & 0.29 & $15.7\pm2.7$ & $118.1\pm0.7$ \\
 &  &  &  & 2460055.5506863 & 117.24 & 0.50 &  & \\
 &  &  &  & 2460113.36141573 & 118.35 & 0.77 &  & \\
HD 136597 & BHB & 4 & 2 & 2478720.15637366 & 34.42 & 0.33 & $22.9\pm2.6$ & $-34.2\pm0.6$ \\
 &  &  &  & 2481034.93857934 & -34.06 & 0.26 &  & \\
 &  &  &  & 2459771.39001661 & -35.73 & 0.51 &  & \\
 &  &  &  & 2460021.63703808 & -35.03 & 1.20 &  & \\
 &  &  &  & 2460218.2364982 & -36.15 & 3.96 &  & \\
 &  &  &  & 2460330.67949533 & -33.71 & 0.34 &  & \\
HD 139459 & BHB & 2 & 0 & 2459771.43189228 & -154.06 & 0.63 & $36.5\pm9.3$ & $-146.3\pm1.8$ \\
 &  &  &  & 2460068.4570065 & noisy & - &  & \\
HD 14829 & BHB & 4 & 2 & 2487894.10188532 & -161.05 & 8.35 & $24.3\pm1.8$ & $-169.7\pm0.9$ \\
 &  &  &  & 2456400.68348332 & -169.28 & 0.59 &  & \\
 &  &  &  & 2459980.20441835 & noisy & - &  & \\
 &  &  &  & 2460219,56570758 & -170.98 & 2.87 &  & \\
 &  &  &  & 2460246.44020835 & noisy & - &  & \\
 &  &  &  & 2460329.22745011 & -169.86 & 0.32 &  & \\
HD 156758 & BHB & 2 & 1 & 2484420.11892656 & -172.18 & 0.44 & $26.3\pm1.9$ & $-172.4\pm0.9$ \\
 &  &  &  & 2459739.53277285 & -172.66 & 0.58 &  & \\
 &  &  &  & 2460088.47570225 & -172.31 & 0.94 &  & \\
HD 167105 & BHB & 3 & 0 & 2459465.40116488 & -173.52 & 0.25 & $21.5\pm0.7$ & $-173.6\pm0.6$ \\
 &  &  &  & 2459471.35834742 & -173.90 & 0.73 &  & \\
 &  &  &  & 2460054.49834952 & -173.85 & 0.29 &  & \\
HD 203563 & MS & 4 & 2 & 2455659.64011932 & -99.39 & 0.66 & $17.7\pm8.2$ & $-100.4\pm0.6$ \\
 &  &  &  & 2460560.50368244 & -102.42 & 0.61 &  & \\
 &  &  &  & 2460560.50368244 & -100.60 & 0.47 &  & \\
 &  &  &  & 2489817.03649306 & -100.08 & 0.44 &  & \\
 &  &  &  & 2459826.37929281 & -100.12 & 0.42 &  & \\
 &  &  &  & 2459827.36878775 & -99.97 & 1.99 &  & \\
HD 208419 & BHB & 1 & 1 & 2492462.64581980 & -27.28 & 0.36 & $28.1\pm4.1$ & $-27.3\pm1.1$ \\
 &  &  &  & 2459772.55875643 & -30.28 & 5.49 &  & \\
HD 213147B & BHB & 3 & 2 & 2488758.58138115 & -211.84 & 0.82 & $19.8\pm1.9$ & $-212.0\pm1.5$ \\
 &  &  &  & 2489720.69988263 & -212.25 & 0.79 &  & \\
 &  &  &  & 2459803.61805032 & noisy & - &  & \\
 &  &  &  & 2460195.45341315 & -211.50 & 1.77 &  & \\
 &  &  &  & 2460218.33586807 & -212.11 & 2.93 &  & \\
HD 231970 & MS & 2 & 0 & 2459821.4125425 & -1.84 & 7.75 & $76.9\pm2.4$ & $-5.1\pm1.8$ \\
 &  &  &  & 2460197.50465654 & -5.31 & 1.81 &  & \\
HD 252940 & BHB & 3 & 1 & 2459286,42855627 & 160.39 & 0.54 & $27.5\pm0.9$ & $160.6\pm0.6$ \\
 &  &  &  & 2459615.341777 & 159.71 & 1.94 &  & \\
 &  &  &  & 2459697.29562812 & 160.44 & 0.37 &  & \\
 &  &  &  & 2460020.30209032 & 160.67 & 0.22 &  & \\
HD 278196 & BHB & 4 & 0 & 2459980.36597813 & -85.80 & 1.73 & $32.5\pm1.8$ & $-85.6\pm2.7$ \\
 &  &  &  & 2460012.31841749 & -85.10 & 1.22 &  & \\
 &  &  &  & 2460042.35724852 & -89.81 & 3.73 &  & \\
 &  &  &  & 2460050.32909249 & -85.57 & 1.81 &  & \\
HD 2857 & BHB & 3 & 3 & 2457087.35906521 & -151.96 & 2.16 & $27.9\pm0.5$ & $-153.6\pm0.6$ \\
 &  &  &  & 2486796.98853002 & -154.06 & 0.61 &  & \\
 &  &  &  & 2487197.36405084 & 149.06 & 5.17 &  & \\
 &  &  &  & 2460195.50067578 & -154.28 & 0.25 &  & \\
 &  &  &  & 2460219.40889409 & -152.78 & 0.83 &  & \\
 &  &  &  & 2460330.19481984 & -153.54 & 0.91 &  & \\
HD 339368 & BHB & 2 & 0 & 2460140.43947607 & -104.32 & 0.67 & $27.3\pm2.8$ & $-105.5\pm1.2$ \\
 &  &  &  & 2460219.29223491 & -105.91 & 0.40 &  & \\
HD 340883 & MS & 1 & 0 & 2460115.39214639 & 4.22 & 2.44 & $332.8\pm4.3$ & $4.2\pm2.4$ \\
HD 3513 & BHB & 2 & 1 & 2492881.45661478 & 4.24 & 0.46 & $28.0\pm1.6$ & $4.6\pm0.9$ \\
 &  &  &  & 2459594.22522236 & 4.83 & 0.62 &  & \\
 &  &  &  & 2460219.45177586 & 5.80 & 1.22 &  & \\
HD 60778 & BHB & 4 & 2 & 2460802.39836298 & 41.67 & 1.10 & $18.8\pm1.1$ & $42.3\pm0.6$ \\
 &  &  &  & 2461490.88151234 & 40.70 & 2.70 &  & \\
 &  &  &  & 2459615.36902492 & 42.37 & 0.40 &  & \\
 &  &  &  & 2459641.3188198 & 42.54 & 0.38 &  & \\
 &  &  &  & 2460061.28364362 & noisy & - &  & \\
 &  &  &  & 2460295.48610832 & 42.25 & 0.33 &  & \\
HD 74721 & BHB & 2 & 2 & 2461699.93178388 & 31.62 & 1.41 & $21.2\pm2.3$ & $31.5\pm1.5$ \\
 &  &  &  & 2462388.20860895 & 30.92 & 1.19 &  & \\
 &  &  &  & 2459615.43185376 & 33.06 & 1.20 &  & \\
 &  &  &  & 2459641.37992292 & 31.28 & 0.59 &  & \\
HD 8376 & BHB & 3 & 0 & 2459467.43017217 & 143.94 & 0.83 & $17.8\pm1.1$ & $144.4\pm1.2$ \\
 &  &  &  & 2459470.41406174 & 144.52 & 0.49 &  & \\
 &  &  &  & 2459825.58155707 & 144.08 & 2.93 &  & \\
HD 86986 & BHB & 3 & 1 & 2465710.51735149 & 10.34 & 0.20 & $16.1\pm1.8$ & $10.4\pm0.3$ \\
 &  &  &  & 2459662.30518544 & 10.36 & 0.20 &  & \\
 &  &  &  & 2459699.41764118 & 10.89 & 0.96 &  & \\
 &  &  &  & 2460012.36664322 & 10.35 & 0.24 &  & \\
HD 87047 & BHB & 3 & 0 & 2459662.34038489 & 137.29 & 0.34 & $21.7\pm2.6$ & $136.9\pm0.9$ \\
 &  &  &  & 2459980.4115668 & 135.62 & 0.67 &  & \\
 &  &  &  & 2459981.56494872 & 134.72 & 16.44 &  & \\
HD 87112 & BHB & 3 & 0 & 2459662.46348934 & -172.78 & 0.43 & $21.2\pm1.6$ & $-172.6\pm0.9$ \\
 &  &  &  & 2459663.31441937 & -172.43 & 0.35 &  & \\
 &  &  &  & 2460068.4180197 & -172.02 & 5.15 &  & \\
HD 93329 & BHB & 2 & 4 & 2473915.13079068 & 205.81 & 1.03 & $21.5\pm2.2$ & $204.9\pm0.6$ \\
 &  &  &  & 2471342.48497296 & 205.03 & 0.37 &  & \\
 &  &  &  & 2470865.56973439 & 205.33 & 0.44 &  & \\
 &  &  &  & 2469688.92063544 & 205.25 & 0.31 &  & \\
 &  &  &  & 2459642.52150551 & 204.48 & 0.32 &  & \\
 &  &  &  & 2459981.53983535 & 204.57 & 0.31 &  & \\
TYC 1093-2946-1 & BHB & 3 & 0 & 2459785.53384058 & -85.96 & 3.14 & $36.1\pm1.5$ & $-85.7\pm2.1$ \\
 &  &  &  & 2459827.4119086 & -85.43 & 0.74 &  & \\
 &  &  &  & 2460208.40478325 & -87.51 & 1.90 &  & \\
TYC 1097-1324-1 & BHB & 3 & 1 & 2489750.19621568 & -207.71 & 0.20 & $23.2\pm1.4$ & $-208.0\pm0.6$ \\
 &  &  &  & 2459742.47382176 & -209.51 & 0.62 &  & \\
 &  &  &  & 2459749.53813884 & -208.98 & 0.65 &  & \\
 &  &  &  & 2459827.45562138 & -209.32 & 0.74 &  & \\
TYC 1342-719-1 & BHB & 3 & 0 & 2459650.28449275 & 53.30 & 1.94 & $25.1\pm9.8$ & $9.1\pm3.9$ \\
 &  &  &  & 2459699.29598315 & -25.20 & 1.90 &  & \\
 &  &  &  & 2460050.27993827 & -24.98 & 3.92 &  & \\
TYC 144-2049-1 & BHB & 2 & 1 & 2467041.84701202 & 194.22 & 0.49 & $19.0\pm1.2$ & $194.1\pm0.9$ \\
 &  &  &  & 2459641.28517679 & 193.95 & 0.42 &  & \\
 &  &  &  & 2459947.42934732 & 194.36 & 0.65 &  & \\
TYC 1738-745-1 & BHB & 3 & 0 & 2459470.37037102 & -155.64 & 0.50 & $20.5\pm0.8$ & $155.0\pm0.6$ \\
 &  &  &  & 2459471.44901513 & -155.16 & 1.46 &  & \\
 &  &  &  & 2459825.53748813 & -154.71 & 0.28 &  & \\
TYC 1914-687-1 & BHB & 3 & 1 & 2472652.99114767 & -67.25 & 0.85 & $38.8\pm1.4$ & $-67.8\pm2.4$ \\
 &  &  &  & 2459650.36749945 & -34.13 & 0.98 &  & \\
 &  &  &  & 2459697.32214202 & -69.75 & 1.71 &  & \\
 &  &  &  & 2459980.30245962 & -48.48 & 1.26 &  & \\
TYC 1916-1763-1 & BHB & 2 & 1 & 2471023.85725328 & -130.91 & 0.41 & $23.6\pm2.1$ & $-131.0\pm0.6$ \\
 &  &  &  & 2459615.39430921 & -134.09 & 1.15 &  & \\
 &  &  &  & 2459641.343063 & -130.91 & 0.29 &  & \\
TYC 1969-388-1 & MS & 2 & 0 & 2459662.37851277 & 96.22 & 1.75 & $141.4\pm3.9$ & $99.3\pm4.2$ \\
 &  &  &  & 2459663.35358007 & 105.20 & 2.42 &  & \\
TYC 1977-1078-1 & BHB & 3 & 0 & 2459662.50370739 & 94.74 & 0.38 & $26.8\pm1.1$ & $95.0\pm0.9$ \\
 &  &  &  & 2459980.49754662 & 96.00 & noisy &  & \\
 &  &  &  & 2460407.42951679 & 95.80 & 0.78 &  & \\
TYC 2058-494-1 & BHB & 2 & 0 & 2459651.64688085 & -169.35 & 0.41 & $37.4\pm8.8$ & $-140.5\pm1.2$ \\
 &  &  &  & 2460069.59725438 & 10.80 & 0.94 &  & \\
TYC 2213-615-1 & BHB & 3 & 0 & 2459802.43567926 & -332.39 & 0.49 & $19.2\pm1.5$ & $-159.2\pm1.2$ \\
 &  &  &  & 2459824.52134728 & -333.67 & 3.38 &  & \\
 &  &  &  & 2460208.45439155 & 334.02 & 0.82 &  & \\
TYC 2225-153-1 & BHB & 3 & 0 & 2459803.39262932 & -195.30 & 1.22 & $22.9\pm1.9$ & $-194.9\pm0.3$ \\
 &  &  &  & 2459823.4798856 & -195.00 & 0.37 &  & \\
 &  &  &  & 2459824.49766417 & -194.31 & 0.64 &  & \\
TYC 2263-704-1 & MS & 2 & 0 & 2459641.23104707 & noisy & - & - & - \\
 &  &  &  & 2459947.28254684 & noisy & - &  & \\
TYC 241-58-1 & MS & 3 & 0 & 2459641.43659333 & -18.70 & 1.88 & $47.8\pm2.7$ & $-21.1\pm1.5$ \\
 &  &  &  & 2459697.35802007 & -17.71 & 0.99 &  & \\
 &  &  &  & 2460070.30781289 & -22.44 & 0.57 &  & \\
TYC 2492-700-1 & MS & 5 & 0 & 2459650.44625247 & 103.43 & 1.93 & $259.3\pm1.3$ & $101.7\pm2.4$ \\
 &  &  &  & 2459697.42321018 & 98.76 & 4.40 &  & \\
 &  &  &  & 2460014.2648642 & 145.02 & 9.03 &  & \\
 &  &  &  & 2460296.52587713 & 103.24 & 2.09 &  & \\
 &  &  &  & 2460329.48925334 & 100.40 & 1.06 &  & \\
TYC 256-372-1 & MS & 4 & 0 & 2459662.42196836 & 111.04 & 2.89 & $256.0\pm1.8$ & $106.8\pm3.6$ \\
 &  &  &  & 2459699.38675201 & 115.87 & 17.37 &  & \\
 &  &  &  & 2460296.59391517 & 106.98 & 1.56 &  & \\
 &  &  &  & 2460330.53340636 & 104.45 & 1.94 &  & \\
TYC 2662-18-1 & MS & 3 & 0 & 2459821.45628907 & 46.54 & noisy & $446.6\pm3.1$ & $46.2\pm5.2$ \\
 &  &  &  & 2460195.40706533 & 50.53 & 4.97 &  & \\
 &  &  &  & 2460219.24826119 & 45.84 & 1.46 &  & \\
TYC 2776-1154-1 & BHB & 2 & 0 & 2459804.55069807 & -233.75 & 1.22 & $16.9\pm1.3$ & $-233.8\pm3.6$ \\
 &  &  &  & 2460218.42614769 & -234.24 & 7.48 &  & \\
TYC 2974-2618-1 & BHB & 3 & 0 & 2459650.32857859 & 171.72 & 0.28 & $28.9\pm4.6$ & $171.6\pm0.6$ \\
 &  &  &  & 2459697.39159727 & 171.39 & 0.28 &  & \\
 &  &  &  & 2460407.38819531 & 171.60 & 2.06 &  & \\
TYC 3114-722-1 & BHB & 2 & 0 & 2459697.45660113 & -166.54 & 0.83 & $20.7\pm1.5$ & $-167.1\pm0.9$ \\
 &  &  &  & 2459821.31919759 & -167.14 & 0.33 &  & \\
TYC 3119-661-1 & BHB & 3 & 0 & 2459699.48053393 & -237.19 & 0.75 & $21.9\pm1.4$ & $-237.9\pm0.9$ \\
 &  &  &  & 2459803.34998214 & -237.78 & 0.45 &  & \\
 &  &  &  & 2459827.32196552 & -238.22 & 0.44 &  & \\
TYC 318-268-1 & MS & 2 & 0 & 2459748.394794 & 18.86 & 1.79 & $220.0\pm4.3$ & $19.0\pm5.4$ \\
 &  &  &  & 2460061.41405695 & 22.49 & 9.76 &  & \\
TYC 3281-1488-1 & MS & 2 & 0 & 2459650.23673097 & -96.44 & noisy & $337.6\pm2.1$ & $-96.2\pm25.8$ \\
 &  &  &  & 2459980.24921287 & -96.23 & 8.60 &  & \\
TYC 3411-1396-1 & BHB & 3 & 0 & 2459615.45634223 & -165.35 & 3.06 & $24.8\pm1.5$ & $-163.2\pm0.9$ \\
 &  &  &  & 2459641.4011447 & -163.56 & 0.38 &  & \\
 &  &  &  & 2460068.38353615 & -161.72 & 0.71 &  & \\
TYC 3433-1068-1 & BHB & 3 & 0 & 2459662.26539877 & 88.68 & 0.84 & $21.6\pm2.6$ & $89.0\pm1.2$ \\
 &  &  &  & 2459663.27109928 & 89.34 & 1.24 &  & \\
 &  &  &  & 2460070.3496312 & 89.10 & 0.51 &  & \\
TYC 3548-814-1 & BHB & 3 & 0 & 2459784.4555634 & -358.04 & 0.59 & $23.1\pm1.1$ & $-358.6\pm0.9$ \\
 &  &  &  & 2460115.34308868 & -358.41 & 0.42 &  & \\
 &  &  &  & 2460197.40381188 & -359.37 & 0.55 &  & \\
TYC 3655-1179-1 & MS & 4 & 0 & 2459467.33746233 & -207.88 & 2.09 & $228.1\pm4.2$ & $-203.9\pm1.8$ \\
 &  &  &  & 2459471.5319235 & -207.07 & 2.21 &  & \\
 &  &  &  & 2459824.60930695 & -209.25 & 2.94 &  & \\
 &  &  &  & 2460020.25511777 & -201.18 & 1.13 &  & \\
TYC 370-296-1 & BHB & 2 & 0 & 2459692.4560421 & -54.72 & 0.61 & $23.2\pm3.0$ & $-54.1\pm1.8$ \\
 &  &  &  & 2460408.55282758 & -57.95 & 3.43 &  & \\
TYC 3838-89-1 & BHB & 2 & 0 & 2459651.56661844 & -232.77 & 0.39 & $39.6\pm2.2$ & $-233.0\pm0.9$ \\
 &  &  &  & 2459802.33667379 & -233.53 & 0.69 &  & \\
TYC 3840-995-1 & BHB & 2 & 0 & 2459675.28714904 & -257.98 & 0.29 & $17.6\pm1.4$ & $258.0\pm0.9$ \\
 &  &  &  & 2459692.29318563 & -257.97 & 0.69 &  & \\
TYC 4151-488-1 & BHB & 2 & 0 & 2459650.4833129 & 106.80 & 0.44 & $26.5\pm3.6$ & $106.9\pm1.2$ \\
 &  &  &  & 2460014.34903417 & 107.81 & 2.31 &  & \\
TYC 4182-1593-1 & MS & 2 & 0 & 2459675.33090072 & -78.81 & 1.75 & $292.8\pm2.6$ & $-78.5\pm4.5$ \\
 &  &  &  & 2459692.32227463 & -74.97 & 6.32 &  & \\
TYC 4205-142-1 & BHB & 3 & 0 & 2459826.32420283 & -201.54 & 1.26 & $24.1\pm1.5$ & $-202.5\pm1.5$ \\
 &  &  &  & 2460014.50298095 & -199.19 & 7.76 &  & \\
 &  &  &  & 2460021.54643579 & -202.62 & 0.52 &  & \\
TYC 4214-1141-1 & BHB & 2 & 0 & 2459339.42264148 & -80.96 & 1.04 & $29.7\pm6.3$ & $-79.4\pm1.8$ \\
 &  &  &  & 2459692.40893738 & -78.53 & 0.78 &  & \\
TYC 500-2426-1 & MS & 2 & 0 & 2459782.53542204 & -78.20 & 7.67 & $317.5\pm4.0$ & $-71.5\pm9.9$ \\
 &  &  &  & 2460219.34333869 & -69.93 & 3.72 &  & \\
TYC 5299-1338-1 & BHB & 2 & 0 & 2460196.62838205 & 110.36 & 0.57 & $28.0\pm2.3$ & $110.8\pm0.9$ \\
 &  &  &  & 2460219.60750012 & 110.98 & 0.40 &  & \\
TYC 552-614-1 & MS & 2 & 0 & 2459802.50395171 & 8.50 & 1.81 & $369.0\pm3.7$ & $5.6\pm5.1$ \\
 &  &  &  & 2460218,28816608 & -12.20 & 4.47 &  & \\
TYC 980-464-1 & BHB & 2 & 0 & 2460113.40738939 & 83.52 & 0.36 & $26.5\pm1.4$ & $83.5\pm1.2$ \\
 &  &  &  & 2460246.20548027 & 82.96 & 3.15 &  & \\

\noalign{\smallskip}
\hline\hline
\end{longtable}

\section{Binarity detection probabilities for targets observed in Ond\v{r}ejov}

\begin{longtable}{lcccccc}
\caption{\label{table:B1} Probabilities of not detecting binarity assuming binary partner masses of 0.8\,\msun, 0.5\,\msun\ and 0.3\,\msun\ assuming a maximum system period of 1000 days or 5000 days for 3\,$\sigma$ confidence.}\\
\hline\hline
\noalign{\smallskip}
 & P<1000 days & P<1000 days & P<1000 days & P<5000 days & P<5000 days & P<5000 days \\
Simbad ID & $M_2=0.8 M_{\odot}$ & $M_2=0.5 M_{\odot}$ & $M_2=0.3 M_{\odot}$ & $M_2=0.8 M_{\odot}$ & $M_2=0.5 M_{\odot}$ & $M_2=0.3 M_{\odot}$ \\
\noalign{\smallskip}
\hline
\noalign{\smallskip}
BD-10 946 & 0.085 & 0.132 & 0.213 & 0.149 & 0.220 & 0.337 \\
BD+00 145 & 0.558 & 0.657 & 0.768 & 0.642 & 0.723 & 0.811 \\
BD+01 548 & 0.195 & 0.280 & 0.406 & 0.330 & 0.409 & 0.517 \\
BD+03 4247 & 0.092 & 0.133 & 0.204 & 0.171 & 0.229 & 0.325 \\
BD+07 2360 & 0.401 & 0.492 & 0.622 & 0.495 & 0.582 & 0.692 \\
BD+10 2 & 0.430 & 0.523 & 0.650 & 0.521 & 0.615 & 0.719 \\
BD+15 34 & 0.086 & 0.132 & 0.209 & 0.205 & 0.263 & 0.352 \\
BD+18 3429 & 0.500 & 0.611 & 0.747 & 0.593 & 0.685 & 0.795 \\
BD+20 5391 & 0.000 & 0.000 & 0.000 & 0.000 & 0.000 & 0.000 \\
BD+25 2602 & 0.078 & 0.115 & 0.184 & 0.191 & 0.249 & 0.333 \\
BD+29 2058 & 0.734 & 0.830 & 0.926 & 0.782 & 0.864 & 0.940 \\
BD+30 623 & 0.706 & 0.787 & 0.875 & 0.759 & 0.828 & 0.896 \\
BD+36 3268 & 0.130 & 0.194 & 0.300 & 0.254 & 0.326 & 0.430 \\
BD+38 574B & 0.097 & 0.155 & 0.257 & 0.184 & 0.260 & 0.389 \\
BD+40 5154 & 0.294 & 0.400 & 0.556 & 0.429 & 0.512 & 0.641 \\
BD+42 2309 & 0.598 & 0.692 & 0.800 & 0.675 & 0.755 & 0.836 \\
BD+43 2344 & 0.461 & 0.567 & 0.688 & 0.556 & 0.649 & 0.745 \\
BD+44 2523 & 0.576 & 0.695 & 0.817 & 0.658 & 0.750 & 0.851 \\
HD 101292 & 0.461 & 0.522 & 0.605 & 0.565 & 0.615 & 0.680 \\
HD 104607 & 0.381 & 0.470 & 0.599 & 0.477 & 0.564 & 0.676 \\
HD 10780 & 0.017 & 0.029 & 0.056 & 0.077 & 0.113 & 0.164 \\
HD 109995 & 0.022 & 0.038 & 0.066 & 0.093 & 0.127 & 0.181 \\
HD 110012 & 0.635 & 0.730 & 0.826 & 0.709 & 0.780 & 0.859 \\
HD 122190 & 0.133 & 0.201 & 0.319 & 0.218 & 0.313 & 0.443 \\
HD 132471 & 0.091 & 0.131 & 0.205 & 0.170 & 0.232 & 0.328 \\
HD 136597 & 0.009 & 0.017 & 0.036 & 0.034 & 0.065 & 0.123 \\
HD 139459 & 0.471 & 0.575 & 0.697 & 0.564 & 0.655 & 0.755 \\
HD 14829 & 0.090 & 0.138 & 0.226 & 0.190 & 0.257 & 0.357 \\
HD 156758 & 0.114 & 0.171 & 0.274 & 0.186 & 0.275 & 0.405 \\
HD 167105 & 0.023 & 0.041 & 0.076 & 0.084 & 0.123 & 0.184 \\
HD 203563 & 0.024 & 0.043 & 0.083 & 0.055 & 0.095 & 0.174 \\
HD 208419 & 0.341 & 0.417 & 0.528 & 0.408 & 0.502 & 0.611 \\
HD 211227 & 0.031 & 0.054 & 0.104 & 0.148 & 0.187 & 0.250 \\
HD 213147B & 0.145 & 0.222 & 0.360 & 0.296 & 0.369 & 0.476 \\
HD 231970 & 0.795 & 0.867 & 0.935 & 0.833 & 0.893 & 0.946 \\
HD 252940 & 0.016 & 0.031 & 0.061 & 0.060 & 0.094 & 0.157 \\
HD 26897 & 0.055 & 0.087 & 0.151 & 0.176 & 0.222 & 0.297 \\
HD 278196 & 0.318 & 0.399 & 0.549 & 0.447 & 0.517 & 0.636 \\
HD 2857 & 0.090 & 0.125 & 0.185 & 0.175 & 0.229 & 0.309 \\
HD 339368 & 0.334 & 0.407 & 0.516 & 0.458 & 0.521 & 0.604 \\
HD 3513 & 0.081 & 0.127 & 0.219 & 0.151 & 0.233 & 0.353 \\
HD 60778 & 0.039 & 0.064 & 0.109 & 0.075 & 0.119 & 0.206 \\
HD 74721 & 0.225 & 0.307 & 0.430 & 0.353 & 0.436 & 0.537 \\
HD 8376 & 0.250 & 0.327 & 0.446 & 0.374 & 0.449 & 0.549 \\
HD 86986 & 0.009 & 0.018 & 0.035 & 0.036 & 0.058 & 0.100 \\
HD 87047 & 0.216 & 0.279 & 0.383 & 0.330 & 0.402 & 0.492 \\
HD 87112 & 0.198 & 0.256 & 0.347 & 0.294 & 0.365 & 0.460 \\
HD 93329 & 0.015 & 0.026 & 0.052 & 0.061 & 0.092 & 0.148 \\
TYC 1093-2946-1 & 0.291 & 0.404 & 0.547 & 0.423 & 0.518 & 0.635 \\
TYC 1097-1324-1 & 0.035 & 0.058 & 0.103 & 0.100 & 0.140 & 0.213 \\
TYC 1342-719-1 & 0.534 & 0.654 & 0.784 & 0.620 & 0.719 & 0.826 \\
TYC 144-2049-1 & 0.292 & 0.357 & 0.461 & 0.348 & 0.437 & 0.552 \\
TYC 1738-745-1 & 0.174 & 0.228 & 0.313 & 0.277 & 0.341 & 0.428 \\
TYC 1914-687-1 & 0.330 & 0.453 & 0.599 & 0.456 & 0.554 & 0.674 \\
TYC 1916-1763-1 & 0.101 & 0.144 & 0.215 & 0.177 & 0.239 & 0.335 \\
TYC 1969-388-1 & 0.814 & 0.854 & 0.902 & 0.848 & 0.880 & 0.921 \\
TYC 1977-1078-1 & 0.112 & 0.172 & 0.287 & 0.198 & 0.279 & 0.410 \\
TYC 2058-494-1 & 0.331 & 0.412 & 0.532 & 0.427 & 0.509 & 0.615 \\
TYC 2213-615-1 & 0.192 & 0.264 & 0.385 & 0.321 & 0.397 & 0.502 \\
TYC 2225-153-1 & 0.321 & 0.371 & 0.438 & 0.450 & 0.489 & 0.547 \\
TYC 2263-704-1 & 0.822 & 0.887 & 0.949 & 0.856 & 0.911 & 0.959 \\
TYC 241-58-1 & 0.193 & 0.276 & 0.418 & 0.326 & 0.413 & 0.530 \\
TYC 2492-700-1 & 0.221 & 0.340 & 0.498 & 0.367 & 0.468 & 0.595 \\
TYC 256-372-1 & 0.391 & 0.517 & 0.676 & 0.504 & 0.605 & 0.734 \\
TYC 2662-18-1 & 0.568 & 0.686 & 0.812 & 0.648 & 0.743 & 0.847 \\
TYC 2776-1154-1 & 0.686 & 0.774 & 0.866 & 0.746 & 0.817 & 0.889 \\
TYC 2974-2618-1 & 0.080 & 0.115 & 0.179 & 0.140 & 0.197 & 0.280 \\
TYC 3114-722-1 & 0.283 & 0.362 & 0.470 & 0.420 & 0.481 & 0.569 \\
TYC 3119-661-1 & 0.120 & 0.174 & 0.267 & 0.286 & 0.328 & 0.404 \\
TYC 318-268-1 & 0.794 & 0.868 & 0.936 & 0.832 & 0.893 & 0.947 \\
TYC 3281-1488-1 & 1.000 & 1.000 & 1.000 & 1.000 & 1.000 & 1.000 \\
TYC 3411-1396-1 & 0.145 & 0.204 & 0.301 & 0.261 & 0.332 & 0.427 \\
TYC 3433-1068-1 & 0.280 & 0.359 & 0.471 & 0.388 & 0.469 & 0.573 \\
TYC 3548-814-1 & 0.087 & 0.141 & 0.227 & 0.208 & 0.271 & 0.364 \\
TYC 3655-1179-1 & 0.328 & 0.445 & 0.577 & 0.457 & 0.548 & 0.659 \\
TYC 370-296-1 & 0.480 & 0.568 & 0.687 & 0.563 & 0.650 & 0.747 \\
TYC 3838-89-1 & 0.301 & 0.380 & 0.497 & 0.432 & 0.497 & 0.592 \\
TYC 3840-995-1 & 0.375 & 0.430 & 0.502 & 0.493 & 0.539 & 0.598 \\
TYC 4151-488-1 & 0.368 & 0.453 & 0.582 & 0.465 & 0.549 & 0.661 \\
TYC 4182-1593-1 & 0.777 & 0.858 & 0.929 & 0.821 & 0.882 & 0.942 \\
TYC 4205-142-1 & 0.254 & 0.339 & 0.460 & 0.395 & 0.464 & 0.562 \\
TYC 4214-1141-1 & 0.470 & 0.579 & 0.695 & 0.568 & 0.659 & 0.753 \\
TYC 500-2426-1 & 0.929 & 0.970 & 0.996 & 0.941 & 0.974 & 0.996 \\
TYC 5299-1338-1 & 0.385 & 0.451 & 0.534 & 0.502 & 0.551 & 0.625 \\
TYC 552-614-1 & 0.783 & 0.858 & 0.927 & 0.824 & 0.884 & 0.942 \\
TYC 980-464-1 & 0.310 & 0.393 & 0.508 & 0.437 & 0.508 & 0.602 \\
\noalign{\smallskip}
\hline\hline
\end{longtable}

\section{ADQL query for common proper motion candidates}

The following query can be used in the https://gea.esac.esa.int/archive/ website where the user table 1 contains the \verb!source_id!, \verb!ra!, \verb!ra_error!, \verb!dec!, \verb!dec_error!, \verb!parallax!, \verb!parallax_error!, \verb!pmra!, \verb!pmra_error!, \verb!pmdec!, \verb!pmdec_error!, \verb!phot_g_mean_mag!, and \verb!bp_rp! of the C24 BHB candidates. This results in the  a file containing the \textit{Gaia} DR3 \verb!source_id! for 118 BHB candidates together with the \textit{Gaia} DR3 data for their respective common proper motion partner candidates:

\begin{longtable}{l}
\noalign{\smallskip}
$\verb!SELECT *, DISTANCE(POINT('ICRS', sd.ra, sd.dec), POINT('ICRS', g.ra, g.dec)) AS dist!$ \\
$\verb!FROM user_rculpan.table1 AS sd!$ \\
$\verb!JOIN gaiaedr3.gaia_source AS g!$ \\
$\verb!ON 1=CONTAINS(POINT('ICRS', sd.ra, sd.dec), CIRCLE('ICRS', g.ra, g.dec, 20.0*sd.parallax/3600.))!$ \\
$\verb!WHERE (sd.source_id != g.source_id)!$ \\
$\verb!AND (((sd.pmra - g.pmra)*(sd.pmra - g.pmra)/!$ \\
$\verb!(sd.pmra_error*sd.pmra_error + g.pmra_error*g.pmra_error) +!$ \\
$\verb!(sd.pmdec - g.pmdec)*(sd.pmdec - g.pmdec)/!$ \\
$\verb!(sd.pmdec_error*sd.pmdec_error + g.pmdec_error*g.pmdec_error)) <= 9.0)!$ \\
$\verb!AND ((sd.parallax - g.parallax)*(sd.parallax - g.parallax)/!$ \\
$\verb!(sd.parallax_error*sd.parallax_error + g.parallax_error*g.parallax_error) <= 9.0)!$ \\
\end{longtable}

\end{appendix}

\end{onecolumn}

\end{document}